\documentclass[12pt]{article}
\usepackage[utf8]{inputenc}
\usepackage{geometry}
\usepackage{graphicx}
\usepackage{amssymb}
\usepackage{amsmath}
\usepackage{mathrsfs}
\usepackage{siunitx}
\usepackage{hyperref}
\usepackage[numbers,sort&compress]{natbib}
\usepackage{float}
\usepackage{subfigure}

\usepackage[font={small}]{caption}
\usepackage{amssymb,mathtools}
\providecommand{\keywords}[1]
{
\begin{flushleft}\textit{Keywords:} #1 \end{flushleft}
}

\title{\Huge{\textbf{Edge Wetting: \\ Steady State of Rivulets \\ in Wedges }}}
\author{Nikolai Kubochkin$ \footnote{corresponding author: kubochkin@ttd.tu-darmstadt.de}$ \\ Tatiana Gambaryan-Roisman\footnote{corresponding author: gtatiana@ttd.tu-darmstadt.de} \\
        \small{Institute for Technical Thermodynamics, Technische Universität Darmstadt} \\ \small{Alarich-Weiss-Straße 10,
64287, Darmstadt, Germany} 
}
\date{} 

\begin{document}

\maketitle

\begin{abstract}
The geometry of rough, textured, fractured and porous media is topologically complicated. Those media are commonly represented by bundles of capillary tubes when modeled. However, angle-containing geometries can serve as a more realistic portrayal of their inner structure. A basic element abidingly inherent to all of them is an open wedge-like channel. The classical theory of capillarity ignoring intermolecular interactions implies that liquid entering the wedge must propagate indefinitely along its spine when the liquid-gas interface is concave. The latter is well-known as a Concus-Finn condition. In the present paper, we show that steady-state rivulets violating Concus-Finn condition can be formed in such channels when the surface forces are taken into account. We present a simple model based on the disjoining pressure approach and analyze the shape of the rivulets in the wedges. Besides, we consider a case when the walls of the wedge are soft and can be deformed by the liquid.
\end{abstract}


\section{Introduction}

Most of surfaces, both natural and man-made, are rough or/and contain pores or cracks \cite{Qur2008}. Usually, they are exposed to a vapor or brought in a contact with a liquid \cite{Yang2020}. Thus, investigation of the capillary transport in rough, porous or cracked media is beneficial for various technologies, including, by not limited to, ink-jet printing and ink production \cite{Kim2011}, biophysics \cite{Carreon2017}, and micro- and nanofluidics \cite{Shou2018,Lee2021}. Combating the climate change stimulates development and improving the techniques of the oil and gas recovery \cite{Singh2019} and sequestration of carbon dioxide \cite{Tokunaga2012}, which, in turn, requires the further comprehension of physics of fluid transport in porous and fractured media. 

\begin{figure}[h]
\centering\includegraphics[width=1\linewidth]{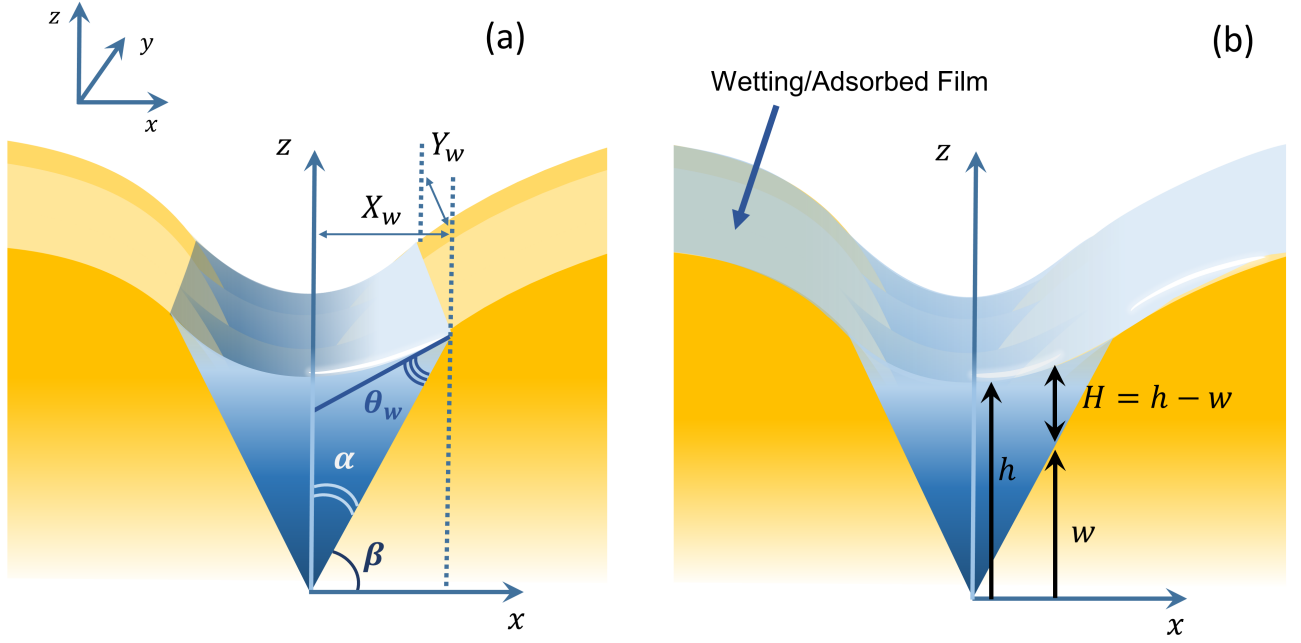}
\caption{Wetting of the symmetrical wedge channel: (a) the wedge wetted by a liquid within the classical capillarity; no adsorbed film is present. $\alpha$ is the half-opening angle of the wedge, $\beta = \pi/2 - \alpha$ is the angle of inclination of the wedge wall, $\theta_w$ is the contact angle, $X_w$ and $Y_w$ are the half-width and length of the wedge, respectively; (b) the wedge wetted by a liquid whose bulk relaxes to the wetting/adsorbed film. Variables $h$ and $w$ measure the distance from the $xy$-plane to the liquid interface and to the wedge wall, respectively.
}
\label{F_Wedge_Picture}
\end{figure}

The geometry of the rough, textured and porous media is generally topologically complicated. In mathematical description, the porous media are commonly and easily represented as bundles of capillary tubes \cite{GambaryanRoisman2014}. However, angle-containing geometries can serve as a more realistic portrayal of inner structure of those media \cite{Tuller1999}. A basic element abidingly inherent to all of them is an open wedge-like channel \cite{Yu2018,ThammannaGurumurthy2018,Weislogel2001,Ponomarenko2011,GambaryanRoisman2009,GambaryanRoisman2019}. 
Statics and dynamics of wetting of the wedges (corners)  pioneered by works of Taylor and \citet{Hauksbee} have been studied for decades. One of the first important insights into the liquid behavior inside the wedge has been reported by \citet{Concus1969}. They showed that the steady shape of the liquid interface is only possible when the liquid-gas interface is convex. Thus, a simple geometrical condition $\alpha + \theta_w > \cfrac{\pi}{2}$, where $\alpha$ is a half of the opening angle of the wedge and $\theta_w$ is the contact angle, with which the liquid meets the wall of the wedge (Figure \ref{F_Wedge_Picture}), should be fulfilled. Otherwise, when the liquid-gas interface is concave, according to the authors \cite{Concus1969}, the solution is unbounded and the capillary pressure gradient drives the rivulet along the corner both in the absence of gravity and, against it, when the it is present. Their result has been augmented by consideration of the wedges with the walls having different wetting properties \cite{Concus1994}. Besides, the maps for stability in the case of more complicated geometries have been plotted in works \cite{Concus1998, Concus2001, Brakke1992, Rascon2016}.  In order to derive the Concus-Finn condition, a simplified approach ignoring the actual non-plane liquid interface shape was developed by \citet{Berthier_General_Condition_2014, Berthier_Whole_Blood_2015}.

The result of Concus and Finn clearly showed the principal difference between physics of the liquid ascending in round capillary tubes and wedges. Indeed, it is well known that if a vertically oriented capillary tube is brought in contact with a liquid pool ($\theta_w < \cfrac{\pi}{2}$), the meniscus starts to climb up over the walls of the tube. The maximal height the meniscus can reach is dictated by the gravity and can be evaluated with the Jurin law \cite{Ponomarenko2011}. Counter-intuitively, if a liquid enters a vertically oriented sharp wedge with the contact angle $\theta_w < \cfrac{\pi}{2} - \alpha$,
gravity cannot stop the ascending rivulet flow, although changes the dynamics of the rivulet propagation. Thus, the rivulet has to propagate infinitely if the wedge is in contact with an infinitely large liquid reservoir. The interface shape in that case is expected to be a hyperbola as was derived in a number of works \cite{Hauskbee1710,Tian2019, Bico2002}. The predicted infinite rise of the rivulet is obviously caused by infinite curvature of the corner, which is not the case for real systems \cite{Bico2002}. Only recently, \citet{Gerlach2020} showed that in the real systems, the finite curvature, determined by the manufacturing process, allows the spreading rivulet to stop. 

The rivulet flows are omnipresent in the porous and rough media, for which  macroscopic considerations of the matter are often insufficient. In order to gain further comprehension of the wetting processes in those media, the nanoscale effects must be incorporated into the models. For the nanoscaled channels, the impact of the surface forces arising from electrostatic, electrodynamic, hydrophobic, steric, and structural interactions is tremendous and cannot be ignored \cite{GambaryanRoisman2014}. Despite the fact that there are a number of works on the wetting of wedges employing the surface force approach, all of them aim to discuss either the critical wetting (filling behavior) \cite{Napirkowski1992,Rejmer1999, Rascn2018, Muotter2017} or the stability of rivulets \cite{GambaryanRoisman2009}. That leads to consideration of a simplified geometry of the problem (two dimensional instead of three dimensional) and no question on the possible rivulet flow arises since the liquid profile is translationally invariant along the channel. The aforementioned works are of undoubted interest; nevertheless, the problems studied do not allow for the answer whether the surface forces stop the rivulet flow in the case when the Concus-Finn condition is violated, and what the shape of such a steady rivulet is. Meanwhile, understanding the limits of the applicability of the Concus-Finn condition can significantly help to understand physics of imbibition, since it is directly related to the extent of the liquid front propagation in porous, textured, or fractured materials.

To the best of our knowledge, the only work discussing the stable shapes of the rivulets in the case when the Concus-Finn condition is violated has been performed by \citet{Wong1992} for the capillaries with the regular polygon cross-sections. It has been shown by them that accounting for the surface forces renders the existence of the steady-states which are not possible within the classical theory of capillarity. However, the disjoining pressure in their work was used primarily as a mathematical tool in order to resolve the contact line problem. 

In the present paper, we study the steady shape of the rivulets in corners/wedges. We demonstrate that introduction of the surface forces embodied in the disjoining pressure leads to the appearance of the steady state of the rivulet in the wedge, although the commonly used Concus-Finn condition is violated. We discuss the influence of the parameters of the disjoining pressure alongside the corner geometry on the steady rivulet profile. Besides, we pay attention to the case when the wedge material is soft and, hence, the wedge walls can be deflected by the traction exerted by the liquid onto the walls of the wedge.
The paper is organized as follows. In section 2, we present the mathematical model and discuss the surface force model. The simulation results for the rivulets in rigid and soft wedges are presented in section 3.

\section{Mathematical model}
\subsection{Governing equations and system parameters}
We consider a concave wedge-shaped channel, which is symmetrical with respect to the $yz$-plane and has a width of $2X_w$ and length of $Y_w$. The walls of the wedge are chemically and physically homogeneous. They are inclined by angle $\beta$ with respect to the $xy$-plane. The opening angle is $2\alpha = \pi - 2\beta$. The variable $h = h(x,y)$ measures the distance between the $xy$-plane and the position of the liquid interface. The variable $H = h - w$, with $w(x) = |x|\tan\beta$ being the profile of the wall, measures the thickness of the liquid layer inside the wedge. The geometry and the orientation of the wedge are shown in Figure \ref{F_Wedge_Picture}. The channel is brought in contact with an infinite reservoir (in practice, it can be a droplet which radius is much larger than the wedge width) located at plane $xz$ the way that a liquid profile $H(x, 0) = I(x)$ at the entrance of the channel is maintained. The wedge is, hence, filled with a liquid of the surface tension $\gamma$. The surrounding gas phase contains the vapor of the liquid so it is favorable for the equilibrium adsorbed/wetting film of thickness $H_{ads}$ to form on the walls of the wedge \cite{Derjaguin1987, Starov2007Book}. Wetting films can be introduced as a tool for elimination of the sharp transition between the liquid-gas interface and solid-gas interface. However, in many cases they are present in real systems and affect their behavior drastically \cite{Guo2020}. The importance of accounting for the wetting films when considering wetting of nanopores has been also reported recently by \citet{Zhang2021}.

For the small angles $\beta$, the interfacial Hamiltonian of the system given can be written within the gradient-squared approximation as 

\begin{equation}
    \mathscr{H}[h(x,y)] = \iint \left ( \frac{\gamma}{2} |\nabla h|^2 + \int_H^{\infty} \Pi(\tilde H)d\tilde H + \Pi\left (H_{ads} \right )H \right ) dxdy,
    \label{E_Hamiltonian}
\end{equation}

\noindent where the first term is so-called Dirichlet energy, $\Pi(H)$ is the disjoining pressure related to the intermolecular interactions inevitably affecting the behavior of the nanoscale systems and the second integral term is the intefacial potential $\mathfrak{S}(H) = \displaystyle\int_H^{\infty} \Pi(\tilde H)d\tilde H$. The last term $\Pi\left (H_{ads} \right )$ is the Lagrange multiplier. We note again that similar to other works on wedges, we assume small $\beta$ allows for assumption that thickness of the adsorbed film $H_{ads}$ in the wedge (measured parallel to $z$-axis as is shown in Figure \ref{F_Wedge_Picture}) is equal to that over the planar wall \cite{Parry1999, Rascn2018}. That leads to the error in the evaluation of $H_{ads}$ which is smaller than $10\%$, where the error of $10\%$ is reached only in the case of the largest $\beta$ considered in the present paper.

The disjoining pressure is considered to encompass a non-retarded repulsive van der Waals (London, Debye, Keesom) term and an attractive electrostatic term. It reads $\Pi(H) = \cfrac{A}{H^3} - K e^{-H/\chi}$ with $A, K$ and $\chi$ being the effective Hamaker constant, strength of electrostatic, and Debay-Hückel, respectively. The Hamaker constant reads as $A_h = 6\pi A$. For the sake of simplicity, we assume that the electrostatic component of the disjoining pressure can be taken in a weak overlap approximation \cite{Tokunaga2012}. Using the exponential term also allows for going beyond DVLO theory as it is a fair approximation of the structural (hydration) pressure, which agrees well with the experimental data \cite{Derjagin1982}. In that case, $K$ is the magnitude of the structural disjoining pressure component and $\chi$ is the hydration layer thickness \cite{Gielok2017}. The chosen expression for the disjoining pressure allows to obtain an S-shaped isotherm having both alpha and beta-branches \cite{Starov2009} (not to be confused with the inclination angles of the wedge).

If the liquid films are ultrathin ($H < 1 \si{\nm}$), $\gamma$ starts to depend on their thickness \cite{Tadmor2008} and can be written in a simplified form as $\gamma(H) = \displaystyle \cfrac{1}{2}\int_{d_{0}}^H \widehat\Pi(\tilde H) d\tilde H$ with $d_{0}$ being the effective cut-off distance and the disjoining pressure $\widehat\Pi(H)$ resulting from the interactions liquid-vacuum-liquid. The general assumption is that the hydrodynamic framework can still capture the system behavior when the confined region is of $H> 1$ $\si{\nm}$ \cite{Zhang2021,Tokunaga2012}. Therefore, in the following, we assume that the hydrodynamics approach is still valid and that the surface tension $\gamma$ is a constant. 

The minimization of (\ref{E_Hamiltonian}) leads to the Derjaguin equation defining the steady-state shape of the rivulet in the wedge:

\begin{equation}
     \nabla^2 H = \frac{\Pi\left (H_{ads} \right ) - \Pi(H)}{\gamma}.
     \label{E_Main_Equation}
\end{equation}

The following boundary conditions have been set: $\cfrac{\partial H}{\partial x} = -\tan\beta$ at $x = 0$, $H = H_{ads}$ at $x = X_w$, $H = I(x)$ at $y = 0$, and $H = V(x)$ at $y = Y_w$ where the functions $I(x)$ and $V(x)$ for the inlet and limiting profiles, accordingly, are to be defined later. Equation (\ref{E_Main_Equation}) can also be derived from the thin-film equation. We note that setting the wetting film thickness $H_{ads}$ can also be interpreted as exposure of the wedge to the atmosphere of vapor of pressure $p_{ads}$. That will be discussed later as well.

\begin{figure}[h]
\centering\includegraphics[width=1\linewidth]{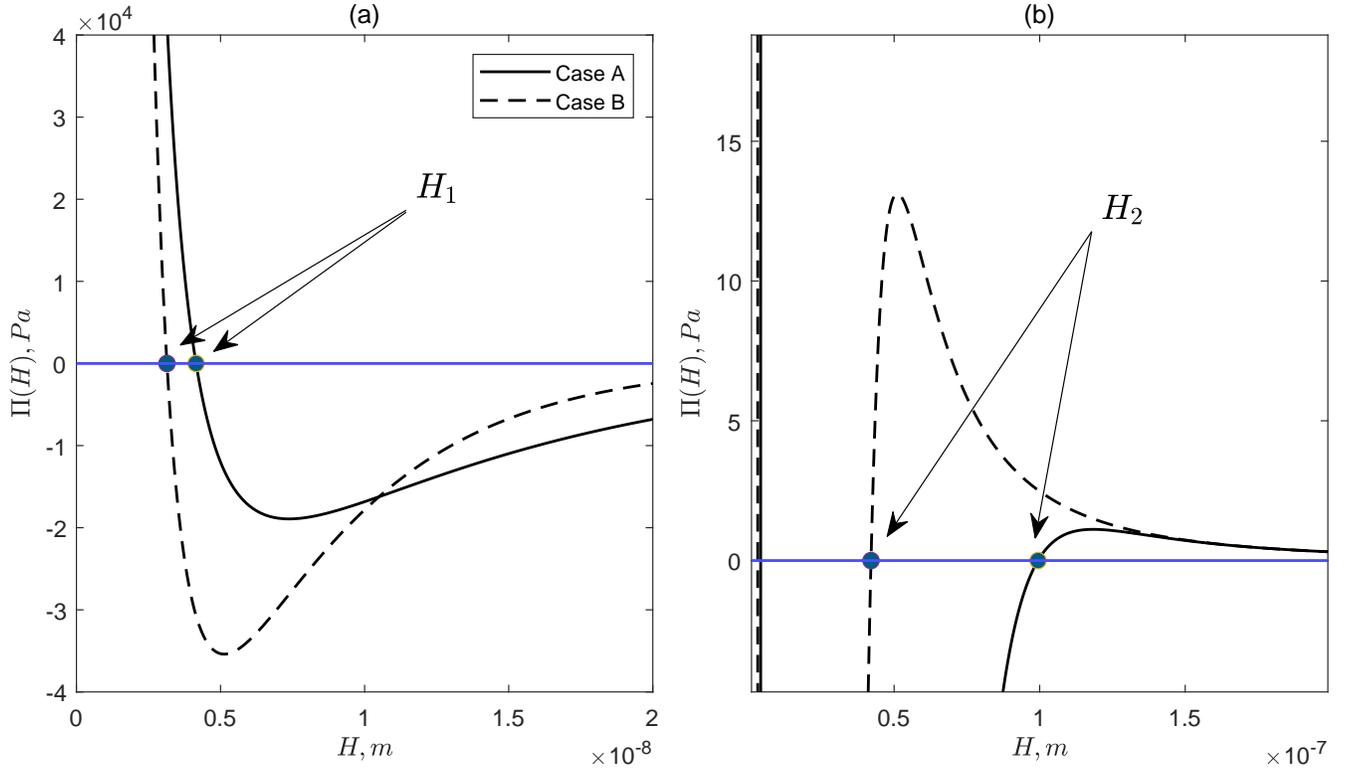}
\caption{Disjoining pressure isotherms used in the present work (Table \ref{T1}): (a) disjoining-conjoining pressure transition defining $H_1$; (b) conjoining-disjoining pressure transition defining $H_2$.
}
\label{F_Isotherms}
\end{figure}

\begin{table}[h]

\centering
\begin{tabular}{l l l l l l l l l}
\hline
\\
System & $A\times 10^{21}, \si{\joule}$ & $K \times 10^{4}, \si{\pascal}$  & $\chi, \si{\nm}$ & $H_1, \si{\nm}$  &$H_2, \si{\nm}$ & $\bar A\times 10^{6}$ & $\bar K\times 10^{2}$  & $\bar \chi\times 10^{2}$ 
\\
\\
\hline
Case A & $2.48$ & $5.25$ & $10.00$ & $4.15$ & $99.43$ & $3.84$ & $7.25$ & $10.06$ \\
Case B & $2.48$ & $15$ & $5.00$ & $3.79$ & $42.10$ & $19.47$ & $8.76$ & $11.89$ \\
\hline

\hline
\end{tabular}
\caption{Parameters of the disjoining pressure isotherms.}
\label{T1}
\end{table}

In the following, we consider a system consisting of an aqueous phase, a solid hydrophilic phase and a gas phase. We assume that changes of the system's parameters are caused mainly by the changes of electrostatic/structural component of the surface force. In practice, that can be reached, for example, by varying the ionic strength of the aqueous phase. As has been reported, the Hamaker constant varies weakly for aqueous solutions and we keep it, therefore, constant. For the liquid and solid interacting across air, $A$ can be evaluated using the Hamaker constants of corresponding liquid (l) and solid (s) media interacting with themselves across the vacuum (v) as $A_h =  \sqrt{A_{h(lvl)}A_{h(svs)}}$ \cite{Israelachvili2011}. We set  $A_{h(lvl)} = 3.73 \times 10^{-20}$J, which corresponds to water \cite{Tokunaga2012, Israelachvili2011}. For the solid substrate, we choose the Hamaker constant to be $A_{h(svs)} \approx 5.86 \times 10^{-20}$J, which finally results in $A_h \approx 4.68 \times 10^{-20}$J or $A \approx 2.48 \times 10^{-21}$J. The disjoining pressure parameters for two cases -- case A and case B -- are presented in Table \ref{T1}. The corresponding disjoining pressure isotherms are shown in Figure \ref{F_Isotherms}(a,b). 
In contrast to the sessile droplets, the van der Waals forces alone still allow for the equilibrium wetting film existence. The difference between cases $A \neq 0$, $K = 0$ and $A \neq 0$, $K \neq 0$ is expected only for relatively thick alpha-films.

Wetting behavior of a system is known to be interrelated with the interfacial potential/disjoining pressure isotherms. Employing the Frumkin-Derjaguin theory \cite{Churaev1995}, the equilibrium contact angle of the liquid meeting the surface covered with the adsorbed film can be evaluated as

\begin{equation}
     \cos{\theta_w} = 1 + \frac{1}{\gamma} \int_{H_{1}}^{\infty} \Pi(H) dH = 1 +  \frac{\mathfrak{S}(H_1)}{\gamma}
\label{E_Angle_Wedge}
\end{equation}

\noindent where $H_{1} = -3\chi \mathcal{W}_0 \left(  -\cfrac{A^{1/3}}{3 K^{1/3} \chi}   \right )$ is the first root of equation $\Pi(H) = 0$ written with the use of the Lambert $\mathcal{W}$-function (see Supplementary Material), and $\mathfrak{S}(H_1)$ defines the minimum of the interfacial potential.  For both isotherms presented, the contact angle is $\theta_w \approx 5^{\circ}$. This contact angle can be observed for a macroscopic meniscus. As can be seen from equation (\ref{E_Angle_Wedge}), keeping the contact angle $\theta_w$ constant means preserving the minimal value of the interfacial potential while allowing for the variations of its shape. The Hamaker constant $A_{h(svs)}$ presented above is a value adjusted to meet the chosen $\theta_w$. Nevertheless, it is still very close to $A_{h(svs)}$ corresponding to mineral surfaces \cite{Tokunaga2012} or a polymer surface such as rubber \cite{Drummond1997}.

Before the solution, equation (\ref{E_Main_Equation}) is written in a dimensionless form using the following dimensionless variables:
$\bar x = x/H_2$, $\bar y = y/H_2$, $\bar H = H/H_2$, $\bar \Pi(\bar H) = \Pi(H)H_2/\gamma$ with $H_2 = -3\chi \mathcal{W}_{-1} \left(  -\cfrac{A^{1/3}}{3 K^{1/3} \chi}  \right )$ being the second root of equation $\Pi(H) = 0$ (see Figure \ref{F_Isotherms}, b). Since generally the amplitude of the attractive (conjoining the interfaces) forces is much stronger than those repulsive (Figure \ref{F_Isotherms}, a-b), choosing $H_2$ for the role of the scaling parameter defining the range of the surface force action is natural. The dimensional width of the wedge was $X_w = 700$ $\si{\nm}$ and was scaled with respect to the $H_2$ of the isotherm chosen. The dimensional length was $Y_w > 500$ $\si{\nm}$ and, analogically, was scaled with respect to the $H_2$ of the isotherm chosen.

\noindent The equation for the equilibrium shape of the rivulet in the dimensionless form is:

\begin{equation}
     \nabla^2 \bar H = \bar\Pi\left (\bar H_{ads} \right ) - \bar\Pi(\bar H).
     \label{E_Main_Equation_Dimensionless}
\end{equation}

The equation (\ref{E_Main_Equation_Dimensionless}) is solved numerically using finite element analysis. It has been checked that the solution converges as maximal element size decreases. It has been checked additionally that the variation of $\bar Y_w$ as well as the number of points, over which the solution has been interpolated, led to only small (less than $2\%$) differences in a decay length $\bar D_{cr}$ (for the definition of $\bar D_{cr}$, see subsection $\ref{SB:Steady_states_of_rivulets})$.

\subsection{Inlet profile and the limiting profile of the rivulet}

As we stated before, we assume that at the inlet $\bar y = 0$, the wedge is connected to the reservoir of infinite volume and the profile of the liquid-gas interface $\bar h(\bar x, 0) = \bar I(\bar x) + |\bar x |\tan \beta$ does not change in time. In real-life systems that can be a nanopore with a macroscopic droplet at its entrance. Assuming that at the entrance, away from the wall of the channel, the shape of the interface $\bar I(\bar x)$ can be obtained from the macroscopic considerations, one can employ the Young-Laplace equation. One integrates it twice and obtain $\bar h(\bar x) = \cfrac{\tan(\beta - \theta_w) (\bar x^2 - \bar X_w^2)}{2\bar X_w} + \bar H_{ads}$. Since the inlet profile must satisfy the boundary conditions and also relax to the wetting film, we introduce a coordinate $\bar x_t$ of the point, where the macroscopic meniscus meets the wetting film with the contact angle $\theta_w$. The inlet profile is, hence, a piece-wise function defined as

\begin{equation}
  \bar h(\bar x) =\begin{cases}
     \cfrac{\tan(\beta - \theta_w)}{\bar x_t} \cfrac{\left (\bar x^2 - \bar x_t^2 \right )}{2} + \bar x_t \tan\beta + \bar H_{ads} + c_1 \bar H_{ads}, & \text{if $\bar x<\bar x_t$}.\\
     \bar H_{ads}, & \text{if $\bar x>\bar x_t + c_2 \left (\bar X_w - \bar x_t \right ) $}.
  \end{cases}
  \label{E_Inlet_Profile}
\end{equation}

\noindent  The value of $\bar x_t$ has been kept equal to $\bar x_t = 0.95\bar X_w$, unless other is stated (see also Supplementary Materials). Real constants $c_1$ and $c_2$ are chosen in a way to allow for a smooth transition from the profile to the wetting film. The constant $c_2$ allowed to change the width of the transition region without changes in $\bar I(0)$. Two regions -- the macroscopic profile and the film -- have been matched (see Supplementary material). Spline interpolation showed overshoots and undulations in the transition region. Therefore, piece-wise cubic Hermite polynomials have been used. Giving less smooth transition zones, it nevertheless allowed to minimize the undesirable waviness there. 

The function $H(\bar x, \bar Y_w) = \bar V(\bar x)$ describes the interface shape far away from the inlet. This function is unknown a priori. Far away from the inlet, the interface is translationally invariant along the $Y$-axis. In that case, the reduced one-dimensional Hamiltonian $\mathscr{H}[h(x)]$ can be written and the liquid profile, hence, can be evaluated using one-dimensional Derjaguin equation 

\begin{equation}
     \frac{d^2 \bar H}{d \bar x^2} = \bar\Pi\left (\bar H_{ads} \right ) - \bar\Pi(\bar H),
     \label{E_Derjaguin_2D}
\end{equation}

\noindent which solution reads

\begin{equation}
     \int_{\bar H_{ads}}^{\bar H} \frac{d \tilde H}{ \sqrt{ \displaystyle2\int_{\bar H_{ads}}^{\tilde H} \left (\bar \Pi\left (\bar H_{ads} \right ) - \bar{\Pi}(\hat{H}) \right ) d \hat{H}}} = \bar x_{ads} - \bar x.
     \label{E_Derjaguin-Solution}
\end{equation}

\noindent We introduce the limiting thickness of the interface at $\bar x = 0$: $\bar H_w = \bar V(0)$. The value of $\bar H_w$ depends on the disjoining pressure and on the slope of the wall as follows from the equation

\begin{equation}
     \frac{\tan^2\beta}{2} = \bar \Pi\left (\bar H_{ads} \right ) \left (\bar H_{w} -  \bar H_{ads} \right ) - \int_{\bar H_{ads}}^{\bar H_w} \left (\bar\Pi(\bar H) \right ) d \bar H.
     \label{E_Middle_Height}
\end{equation}

\noindent As long as the disjoining pressure isotherm contains the exponential tail, the equation (\ref{E_Middle_Height}) is transcendental and the integral in (\ref{E_Derjaguin-Solution}) cannot be taken analytically. Therefore, equations (\ref{E_Derjaguin-Solution}) and (\ref{E_Middle_Height}) were integrated numerically, subject to boundary conditions $\bar H = \bar H_{ads}$ at $\bar x = \bar x_w$ and $\cfrac{d \bar H}{d \bar x} = \tan\beta$ at $\bar x = 0$. In order to avoid a singularity when the adsorbed film thickness is reached, the integration was started from $\bar H_{ads} + \epsilon$ where for $\epsilon$ the condition $\epsilon/\bar H_{ads} \ll 1$ is fulfilled. The function $V(x)$ was obtained as $V(x) = \bar H(\bar x) + |x|\tan\beta$. The solution of (\ref{E_Derjaguin_2D}) has been checked by the direct differentiation as well as by solving it as a boundary value problem fully numerically.

When the wetting film is very thin: $\bar H_{ads} = \phi \bar H_1$ with $\phi \ll 1$, one can expect the limiting height of the rivulet $\bar H_w$ to be smaller than the height $\bar H_{c1} = -4  \bar \chi \mathcal{W}_0 \left(  -\cfrac{ (3\bar A)^{1/4}}{4 \bar K^{1/4} \bar \chi^{3/4}}  \right )$, where the disjoining pressure isotherm finds its minimum. In this case, the disjoining pressure isotherm can be linearized and equation (\ref{E_Middle_Height}) can be easily solved analytically. The limiting height $\bar H_w$ can be obtained as 

\begin{equation}
    \bar H_w = \bar H_{ads} + \tan\beta \sqrt{\frac{\bar H_{1} - \bar H_{ads}}{\bar \Pi\left (\bar H_{ads} \right )}}
    \label{E:Linearized_Equation_Limiting_Height}
\end{equation}

\noindent The derivation and discussion of (\ref{E_Inlet_Profile}) and (\ref{E:Linearized_Equation_Limiting_Height}) are presented in Supplementary Materials.

\section{Results and discussion}
\subsection{Rigid Wedges}
\label{SB:Steady_states_of_rivulets}
\begin{figure}
\centering\includegraphics[width=1\linewidth]{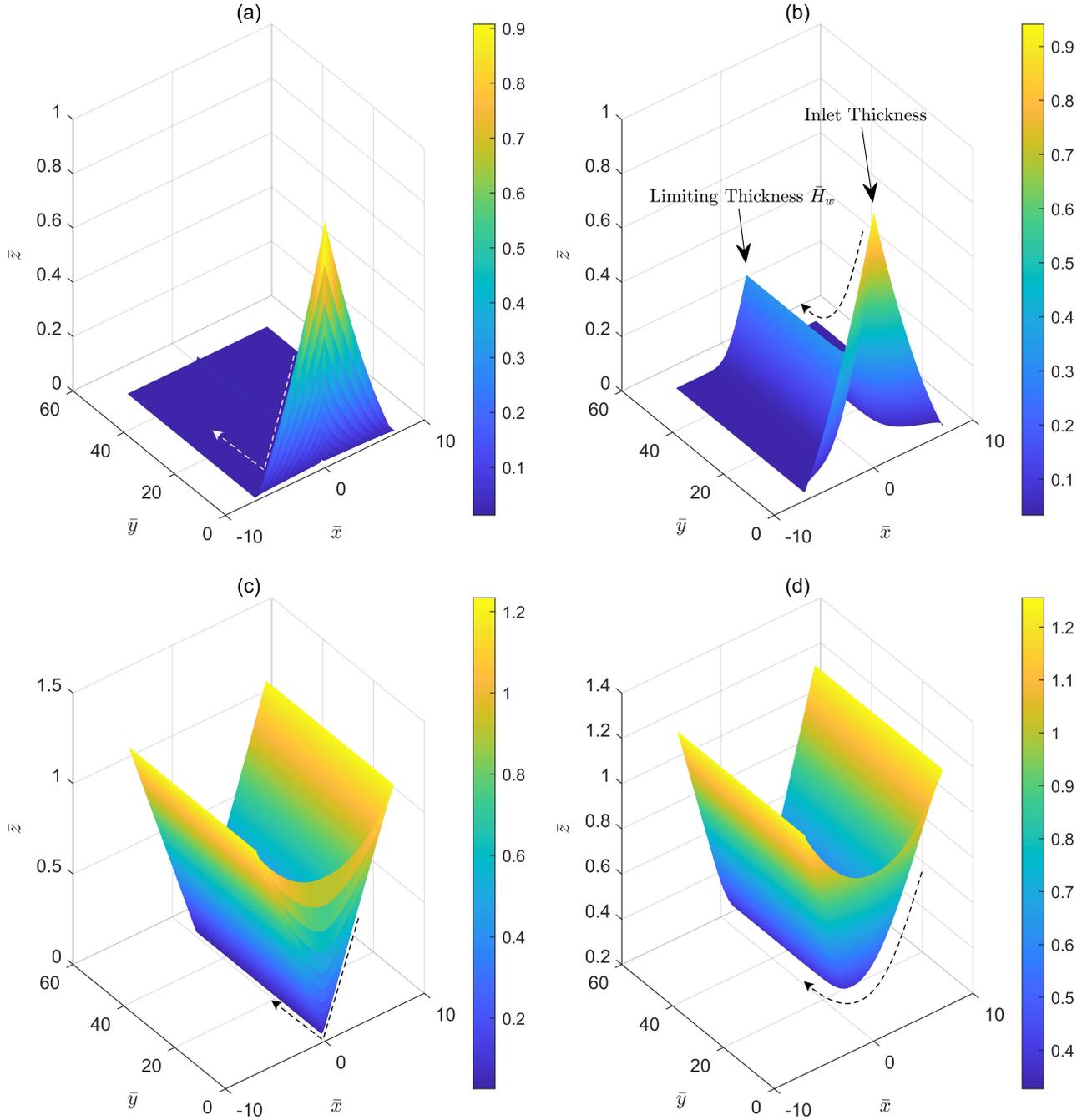}
\caption{Wetting of the symmetrical wedge-shaped channel. The thickness of the liquid $\bar H$ in the channel when the equilibrium adsorbed film thickness (a) $H_{ads} = 0.0125$ ($\phi =0.3$) and (b)  $H_{ads} = 0.0334$ ($\phi =0.8$). The profile of the liquid-gas interface $\bar h$  in the channel when the equilibrium adsorbed film thickness (c) $H_{ads} = 0.0125$ ($\phi =0.3$) and (d)  $H_{ads} = 0.0334$ ($\phi =0.8$). The inclination angle is $\beta = 2\theta_{w}$. The isotherm A has been used for the calculations. The color bars are to show the thickness/height distribution. Less number of points than the solutions contain was used for plotting in order to decrease rendering time and avoid low-level graphic errors.}
\label{F_3DProfiles}
\end{figure}

The presence of the surface forces stabilizing the adsorbed/wetting film allows for the steady state meniscus in the wedge even when the Concus-Finn condition is not fulfilled. The solution of equation (\ref{E_Main_Equation_Dimensionless}) for the adsorbed film thicknesses $\bar H_{ads} = 0.0125$, ($\phi =0.3$) and $\bar H_{ads} = 0.0334$, ($\phi =0.8$) are illustrated in Figure \ref{F_3DProfiles}. The thickness of the liquid layer $\bar H$ and and the liquid-gas interface  $\bar h$ are shown in Figure \ref{F_3DProfiles}(a, b) and (c, d), correspondingly. As is set by the boundary conditions, the thickness of the liquid layer is maximal at the entrance of the wedge.  The liquid layer thickness at the plane $\bar x = 0$ decays until the limiting thickness $\bar H_{w}$ dictated by the solution of equation (\ref{E_Middle_Height}) is reached. One can see that when the adsorbed film is thinner (Figure \ref{F_3DProfiles}, a, c), the thickness of the liquid filling the wedge diminishes down to the limiting profile  $\bar V(\bar x)$ very close to the entrance (schematically shown by the arrows). In contrast to that, when the adsorbed film is thicker (Figure \ref{F_3DProfiles}, b, d), the liquid thickness decreases gradually and reaches the limiting profile $\bar V(\bar x)$ far away from the entrance.

\begin{figure}
\centering\includegraphics[width=1\linewidth]{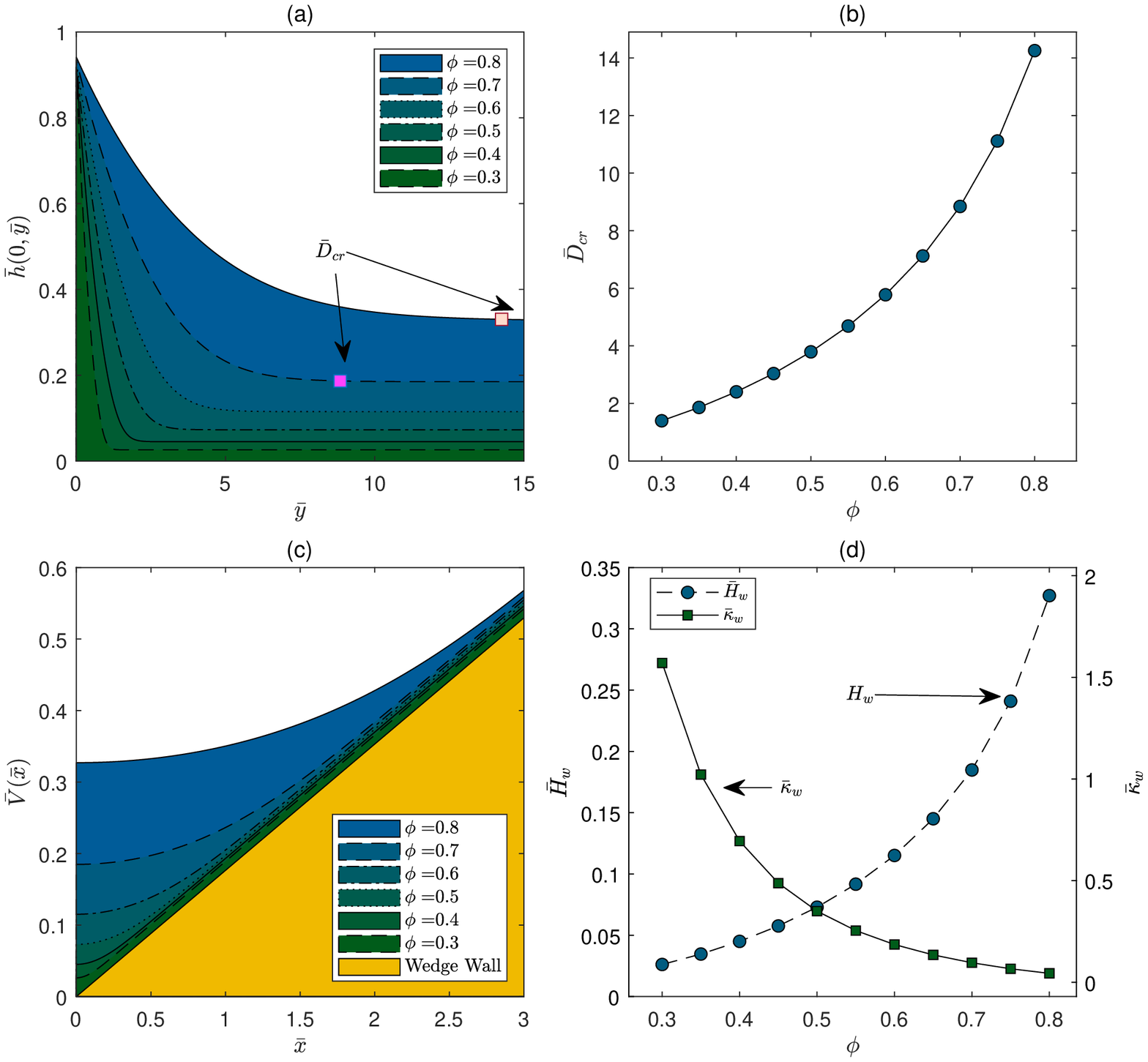}
\caption{Wetting of the wedge channel; (a) thickness of the rivulet at the symmetry axis ($\bar x = 0$) calculated with the different adsorbed film thicknesses $\bar H_{ads}$; (b) dependence of the decay length $\bar D_{cr}$ on the  adsorbed film thickness; the decay lengths are illustrated in (a); (c) limiting profiles $\bar V(\bar x)$ of the rivulets calculated with the different adsorbed film thicknesses $\bar H_{ads}$; (d) dependence of the limiting curvature $\bar \kappa_{w}$ and limiting height $\bar H_{w}$  of the rivulet at the symmetry axis ($\bar x = 0$) on the  adsorbed film thickness (or parameter $\phi$). The inclination angle is $\beta = 2\theta_{w}$. The system behavior is defined by the isotherm A. The lines in (b) and (d) are to guide an eye. }
\label{F_FProfiles}
\end{figure}

\begin{figure}[h]
\centering\includegraphics[width=0.7\linewidth]{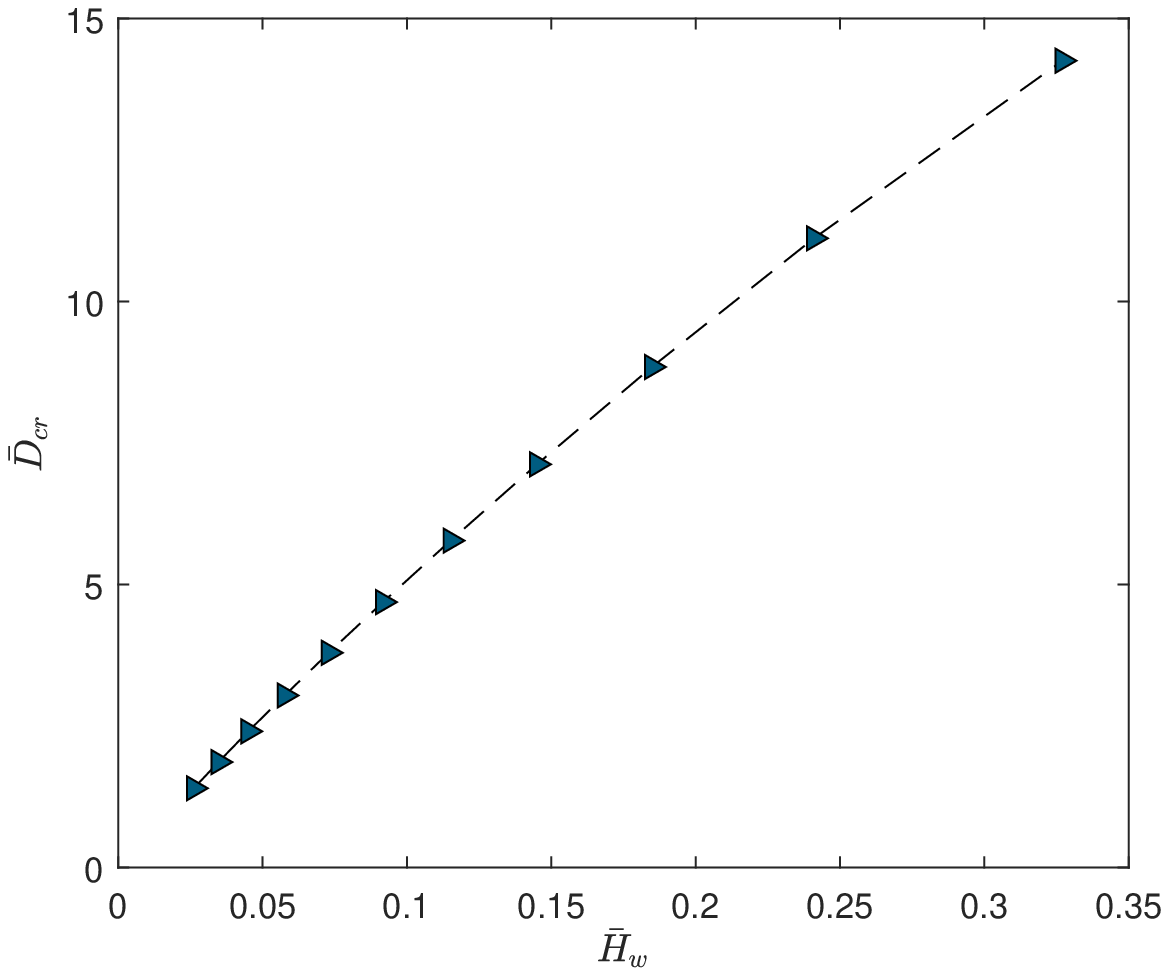}
\caption{Dependence of the decay length $\bar D_{cr}$ on the limiting thickness $\bar H_{w}$ of the profile $\bar V(\bar x)$. The inclination angle is $\beta = 2\theta_{w}$. The system behavior is defined by the isotherm A.}
\label{F_Height_Distance}
\end{figure}

The smoothness of the transition from the inlet to the limiting profile can be illustrated more clear by plotting the thickness $\bar H$ of the liquid layer at $\bar x = 0$ (Figure \ref{F_FProfiles}, a). We quantify this transition by introducing the decay length $\bar D_{cr}$. We define $\bar D_{cr}$ as the length, at which the relative difference between $\bar H(0, \bar y)$ and $\bar H_{w}$ is less than $1\%$. The decay length is shown in Figure \ref{F_FProfiles}(a) by the black arrows and squared markers. The calculated values of $\bar D_{cr}$ are presented in Figure \ref{F_FProfiles}(b). One can see that $\bar D_{cr}$ increases with the increasing ratio $\phi$ (increasing $\bar H_{ads}$). The observed trend is obviously related to changes in the limiting profile $\bar V(\bar x)$ of the rivulet resulting from equation (\ref{E_Derjaguin_2D}). The limiting profile is in turn dictated by interplay of curvature-induced and disjoining pressures. In Figure \ref{F_FProfiles}(c-d), the limiting rivulet profiles as well as the dependency of the dimensionless curvature $\bar \kappa_{w} =\cfrac{\partial^2 \bar H}{\partial \bar x^2}(0,\bar Y_w)$ and dimensionless limiting height of the rivulet $\bar H_{w} = \bar V(0)$ on the ratio $\phi$ are shown. The limiting height increases as the adsorbed film $\bar H_{ads}$ increases, while the opposite trend is observed for the dimensionless curvature: it decreases with the increasing $\bar H_{ads}$. In Figure \ref{F_Height_Distance}, decay length $\bar D_{cr}$ is plotted against the limiting height $\bar H_{w}$. As the adsorbed/wetting film thickness increases causing a rise of $\bar H_{w}$, the decaying length $\bar D_{cr}$ increases as well.

\begin{figure}[h]
\centering\includegraphics[width=0.7\linewidth]{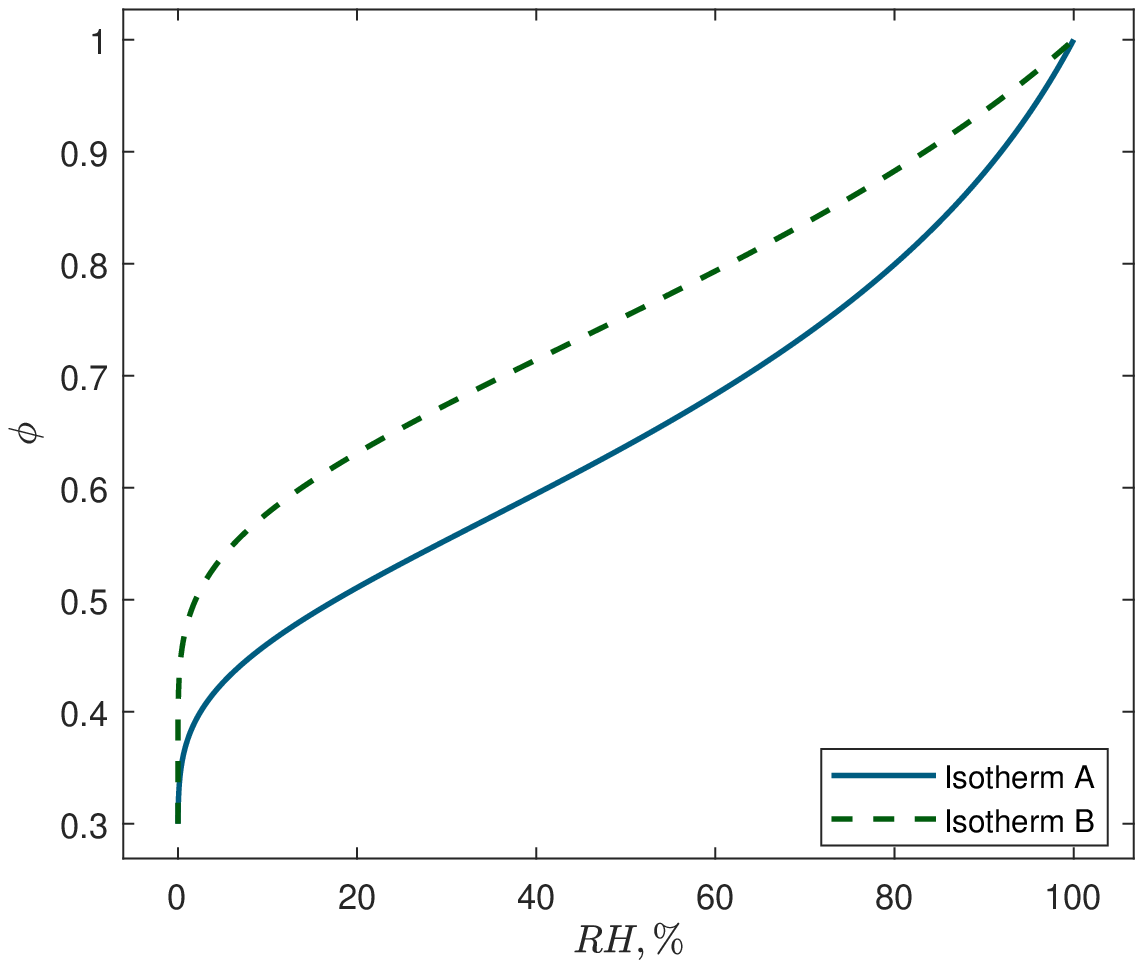}
\caption{Influence of the relative humidity on the thickness of the adsorbed/wetting film at $T = 293$ K ($\nu_m = 18 \times 10^{-3}$ kg/m) for isotherms A and B.}
\label{F_Humidity}
\end{figure}

As has been briefly mentioned before, the adsorbed/wetting film of a constant equilibrium thickness $\bar H_{ads}$ is formed on the walls of the wedge when it is exposed to the vapor atmosphere of pressure $\bar p_{ads}$. The value of the corresponding vapor pressure or relative humidity $RH = \cfrac{p_{ads}}{p_s}$ (where $p_s$ is the pressure of the saturated vapor) can be calculated using Derjaguin-Churaev-M\"uller theory of the surface forces \cite{Derjaguin1987}:

\begin{equation}
    RH = \exp{\left (-\cfrac{\nu_m \Pi(H_{ads})}{RT}\right )}
    \label{Kelvin_eq}
\end{equation}

\noindent where $\nu_m$ is a molar volume of liquid, $R$ is a gas constant, $T$ is temperature. We plot the dependence of $\phi$ on $RH$ in Figure \ref{F_Humidity}. One can see that very thin wetting films can exist even when humidity is close to zero. The isothermal increase of the relative humidity $RH$ leads to the increasing limiting height $\bar H_{w}$ and, therefore, to the increasing decay length $\bar D_{cr}$ of the rivulet.

\begin{figure}
\centering\includegraphics[width=1\linewidth]{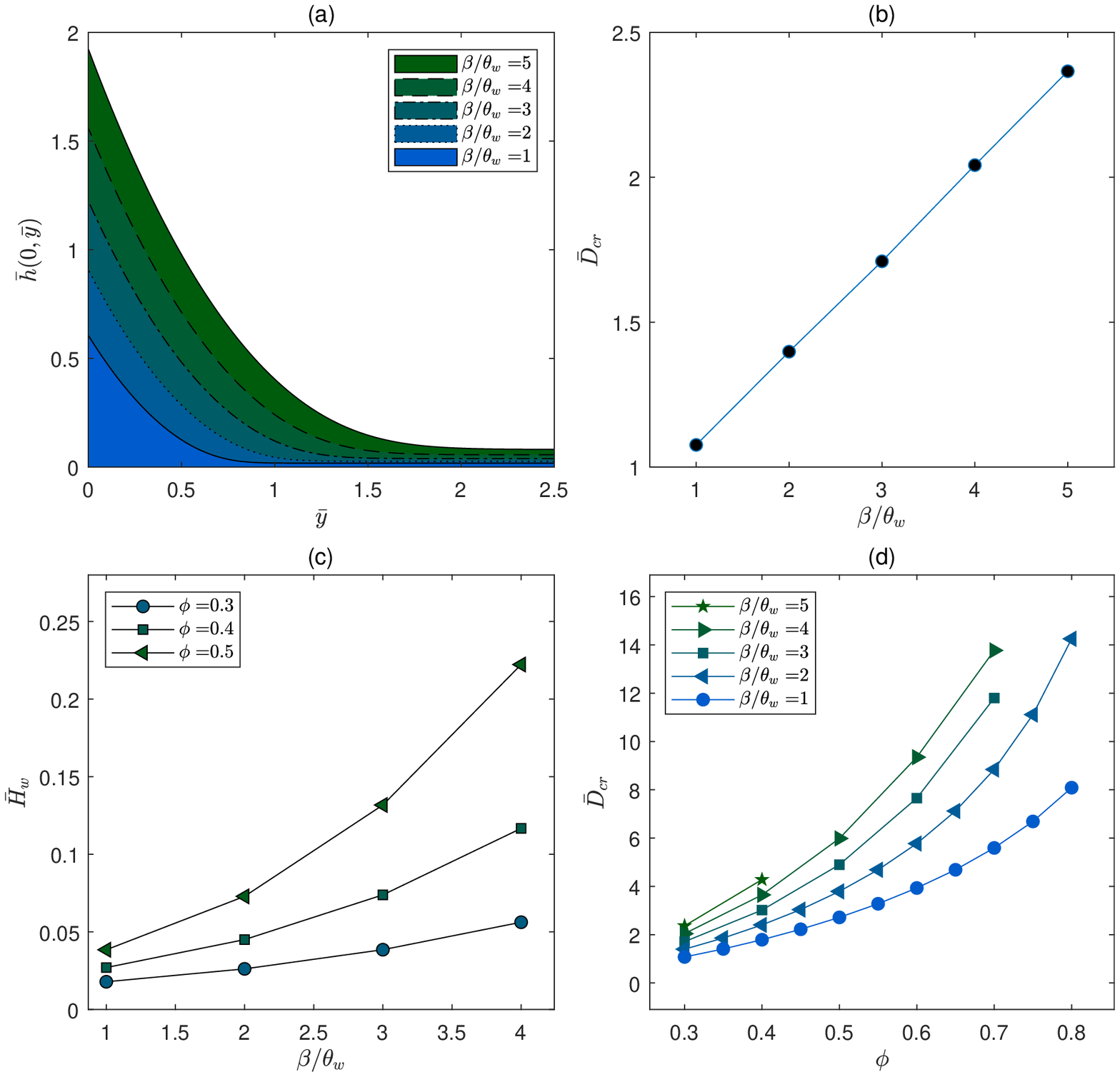}
\caption{Wetting of the wedges with different inclination angle $\beta$: (a) thickness of the liquid layer at the symmetry axis ($\bar x = 0$) calculated with the different inclination angles $\beta$ ($\phi = 0.3$); (b) influence of the inclination angle $\beta$ on the decay length $\bar D_{cr}$; (c) influence of the inclination angle $\beta$ on the on the limiting thickness $\bar H_{w}$ of the profile $\bar V(\bar x)$ for different $\phi$; (d) influence of the ratio $\phi$ (adsorbed film thickness) on the decay length  $\bar D_{cr}$ for wedges with the walls inclined by different angles $\beta$. The system behavior is defined by the isotherm A.}
\label{F_Angle_Dependence}
\end{figure}

The rivulet shape is drastically affected by the geometry of the wedge as well. In order to show that, we solve equation (\ref{E_Main_Equation_Dimensionless}) for five different inclination angles $\beta$ chosen in the way that the steady liquid profile in the sense of the Concus-Finn consideration does not exist. The ratio $\beta/\theta_w$ varies from $1$ to $5$. 

In Figure \ref{F_Angle_Dependence}(a), the liquid profiles corresponding to different inclination angles $\beta$ (ratios $\beta/\theta_w$) are presented. Here, we first consider the case when increasing $\beta$ leads to increasing of the maximal film thickness at the inlet. That naturally follows from inlet condition (\ref{E_Inlet_Profile}) with $\bar x_t$ kept constant. Interestingly, when closing the wedge in a described way, the decay length of the rivulet $\bar D_{cr}$ increases linearly with $\beta$ (Figure \ref{F_Angle_Dependence}, b). In contrast to that, the dependence of the limiting height of the rivulet $\bar H_{w}$ on the ratio of the wedge inclination angle $\beta$ and contact angle $\theta_w$ demonstrates the non-linear growth (Figure \ref{F_Angle_Dependence}, c). The dependencies $\bar H_{w}(\beta)$ are plotted in Figure \ref{F_Angle_Dependence}(c) for different values of $\phi$. It can be seen that when the wetting film is thin, the effect of the the wedge inclination is not strong, while it becomes more pronounced when the adsorbed film is thicker ($RH$ is higher). It is shown in Figure \ref{F_Height_Distance} that the decay length $\bar D_{cr}$ increases with the increasing $\bar H_{w}$. Therefore, one should expect  $\bar D_{cr}$ is more significantly affected by the wedge angle when the adsorbed film is thicker ($RH$ is higher). In Figure \ref{F_Angle_Dependence}(d), the decay length $\bar D_{cr}$ is plotted against $\phi$ for different ratios $\beta/\theta_w$ and the expected trend can be observed. 

\begin{figure}[h]
\centering\includegraphics[width=1\linewidth]{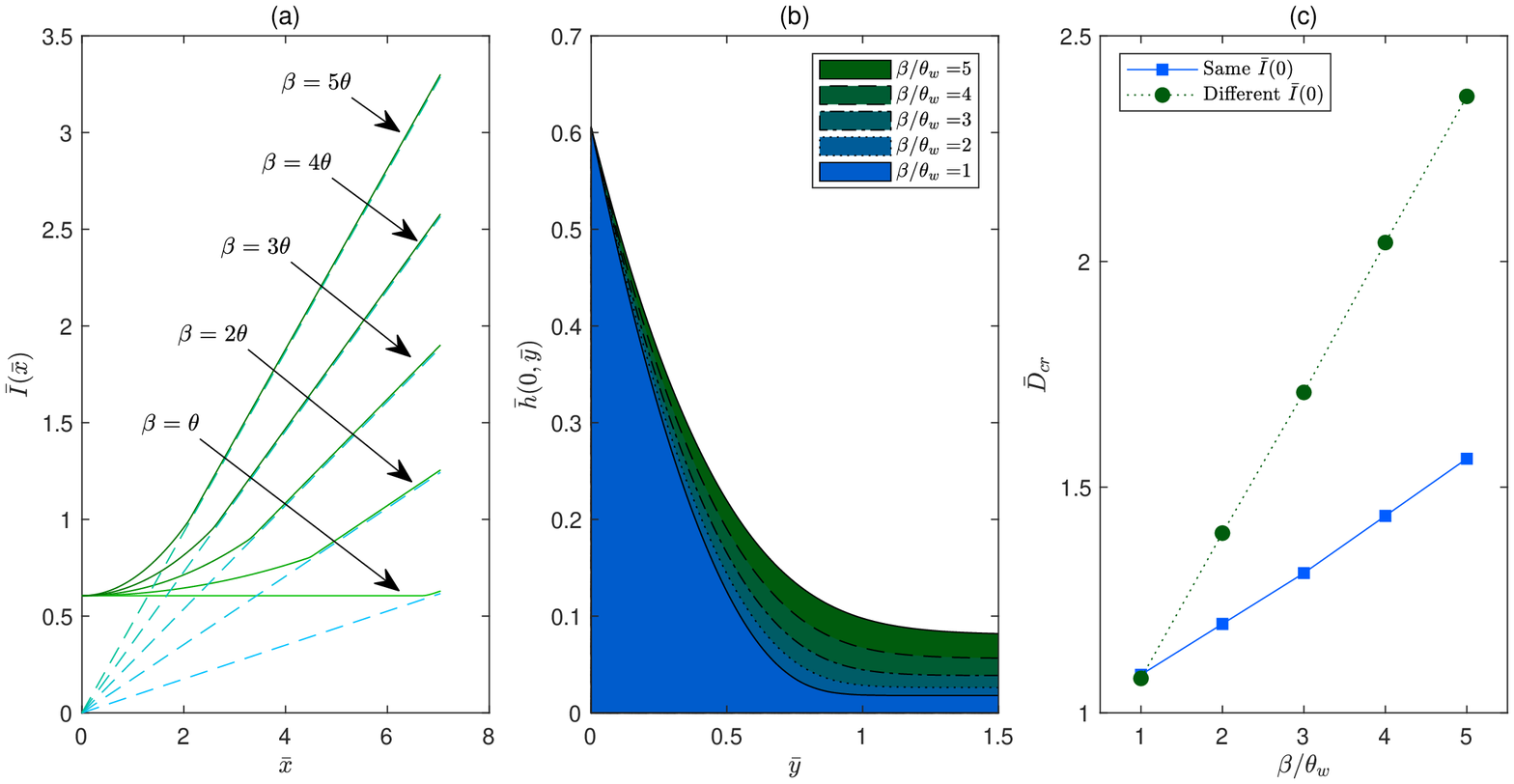}
\caption{Wetting of the wedges with the different inclination angles $\beta$: (a) inlet profiles $\bar I(\bar x)$ for different inclination angles $\beta$ ($\phi = 0.3$). The thickness of the liquid layer at the inlet, at the symmetry axis ($\bar x = 0$) is fixed; (b) thickness of the liquid layer at the symmetry axis ($\bar x = 0$) calculated with the different inclination angles $\beta$ ($\phi = 0.3$); (c) influence of the inclination angle $\beta$ on the decay length $\bar D_{cr}$. The system behavior is defined by the isotherm A.}
\label{F_Angle_Dependence_Same_Inlet_Thickness}
\end{figure}

As has been stated afore and shown in Figure \ref{F_Angle_Dependence}(a), the thickness of the liquid layer at the center of the inlet $\bar I(0)$ has not been preserved when closing the wedge (increasing wedge angle $\beta$). The variation of $\bar I(0)$ can affect the dependence $\bar D_{cr}(\beta)$. In order to eliminate the effect of $\bar I(0)$ on $D_{cr}(\beta)$, we perform calculations, in which $\beta$ has been varied, while $\bar I(0)$ is kept constant. The latter was possible by varying the value of $\bar x_t$ from $\approx 0.29\bar X_w$ ($\beta = 5\theta$) to $\approx 0.95\bar X_w$ ($\beta = \theta$). The values of $\bar x_t$ as well as of the parameters $c_1$ and $c_2$ of the inlet profiles are given in Supplementary materials. The inlet profiles $\bar I(\bar x)$ employed are shown in Figure \ref{F_Angle_Dependence_Same_Inlet_Thickness}(a). The resulting thicknesses of the rivulets along the symmetry axis of the wedge are presented in Figure \ref{F_Angle_Dependence_Same_Inlet_Thickness}(b). Despite the fact that $\bar I(0)$ is now the same, the rivulet in the wedge, where the wetting/adsorbed film is thicker, still extends over the longer distance. The dependencies of $\bar D_{cr}$ on $\beta/\theta_w$ for the cases of both same and different $\bar I(0)$ are shown in Figure \ref{F_Angle_Dependence_Same_Inlet_Thickness}(c). Keeping thickness of the rivulet at the inlet $\bar I(0)$ constant does not break the linear dependence of $\bar D_{cr}$ on $\beta/\theta_w$, although makes it less pronounced.

\begin{figure}
\centering\includegraphics[width=0.7\linewidth]{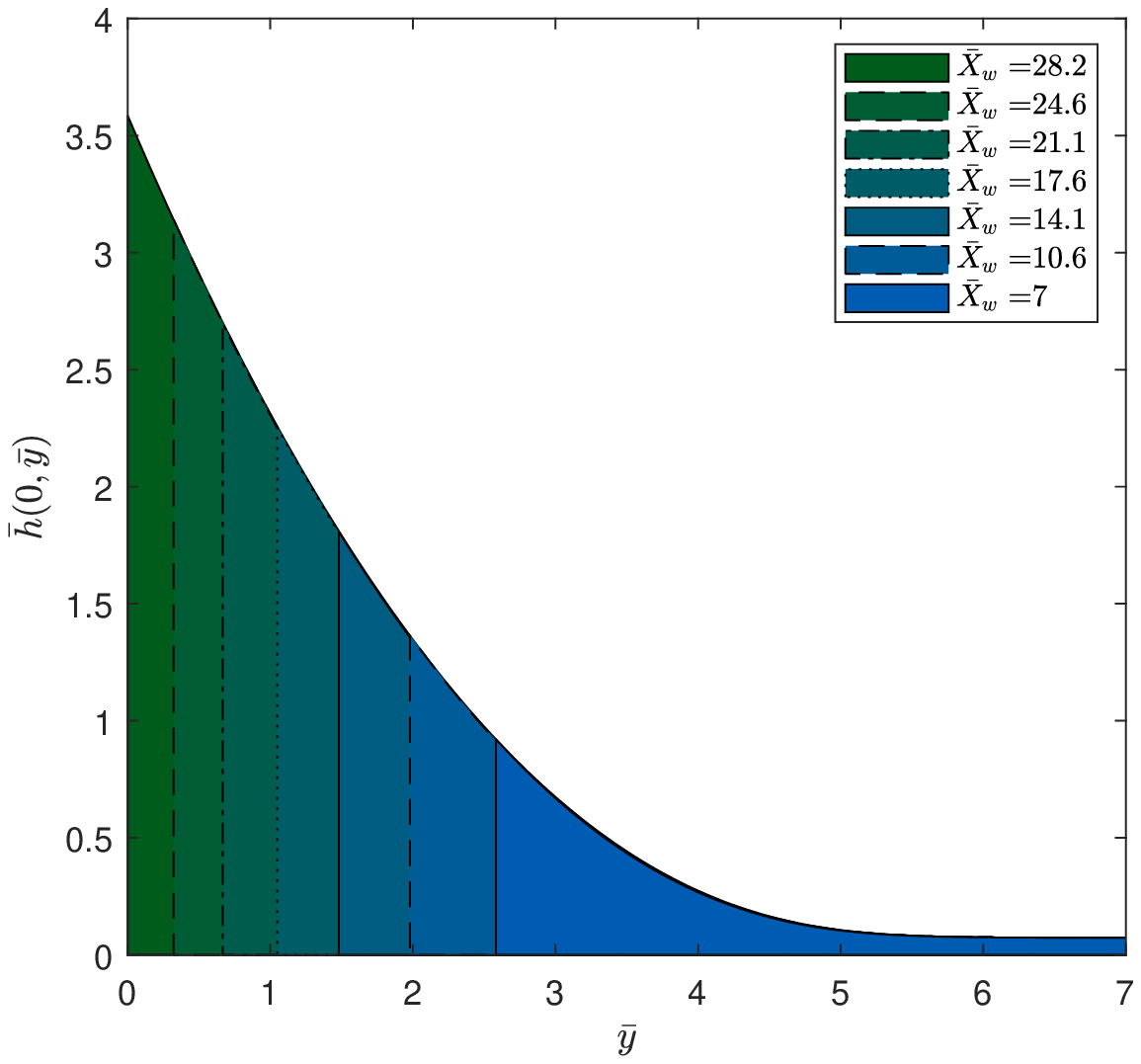}
\caption{Wetting of the wedges of different size: Thickness of the rivulets at the symmetry axis ($\bar x = 0$). The inclination angle is $\beta = 2\theta_{w}$. The system behavior is defined by the isotherm A.}
\label{F_Wedge_Size_Influence}
\end{figure}

According to the Washburn law, the extent of the liquid front in the porous media is related to the size of the pores \cite{GambaryanRoisman2014}. Moreover, the kinetics of the liquid front in corner geometry depends on the meniscus height at the inlet \cite{GambaryanRoisman2019}. Therefore, it is also of interest to study how the size of the wedge affects the decay length $\bar D_{cr}$ of the static rivulet. The width of the wedge was varied from $\bar X_w \approx 7$ to $\bar X_w \approx 28$, the length of the wedge $\bar Y_w$, the adsorbed film thickness $\bar H_{ads}$, and the wedge inclination angle $\beta$ were kept constant. The profiles of the liquid $\bar h(\bar y)$ filling corresponding wedges are presented in Figure \ref{F_Wedge_Size_Influence}. In the larger wedges, the liquid thickness at the inlet is larger and the rivulet extends along a larger distance. Therefore, the decay length $\bar D_{cr}$ increases with increasing size of the wedge. All the rivulets have been initially computed with the inlet located at $\bar y = 0$. However, the shape of the profiles can give us an idea that $\bar h(\bar y)$ being plotted and shifted along $\bar y$ with respect to the profile of the liquid in the widest wedge might match. Indeed, $\bar h(\bar y)$ for the wedges with the widths of up to fourfold difference coincide almost perfectly when plotted with a subsequent shift (Figure \ref{F_Wedge_Size_Influence}). That observation can be of significant help when a profile of a liquid rivulet is known for a large angular pore and the length of a smaller rivulet is to be evaluated. 

\begin{figure}
\centering\includegraphics[width=1\linewidth]{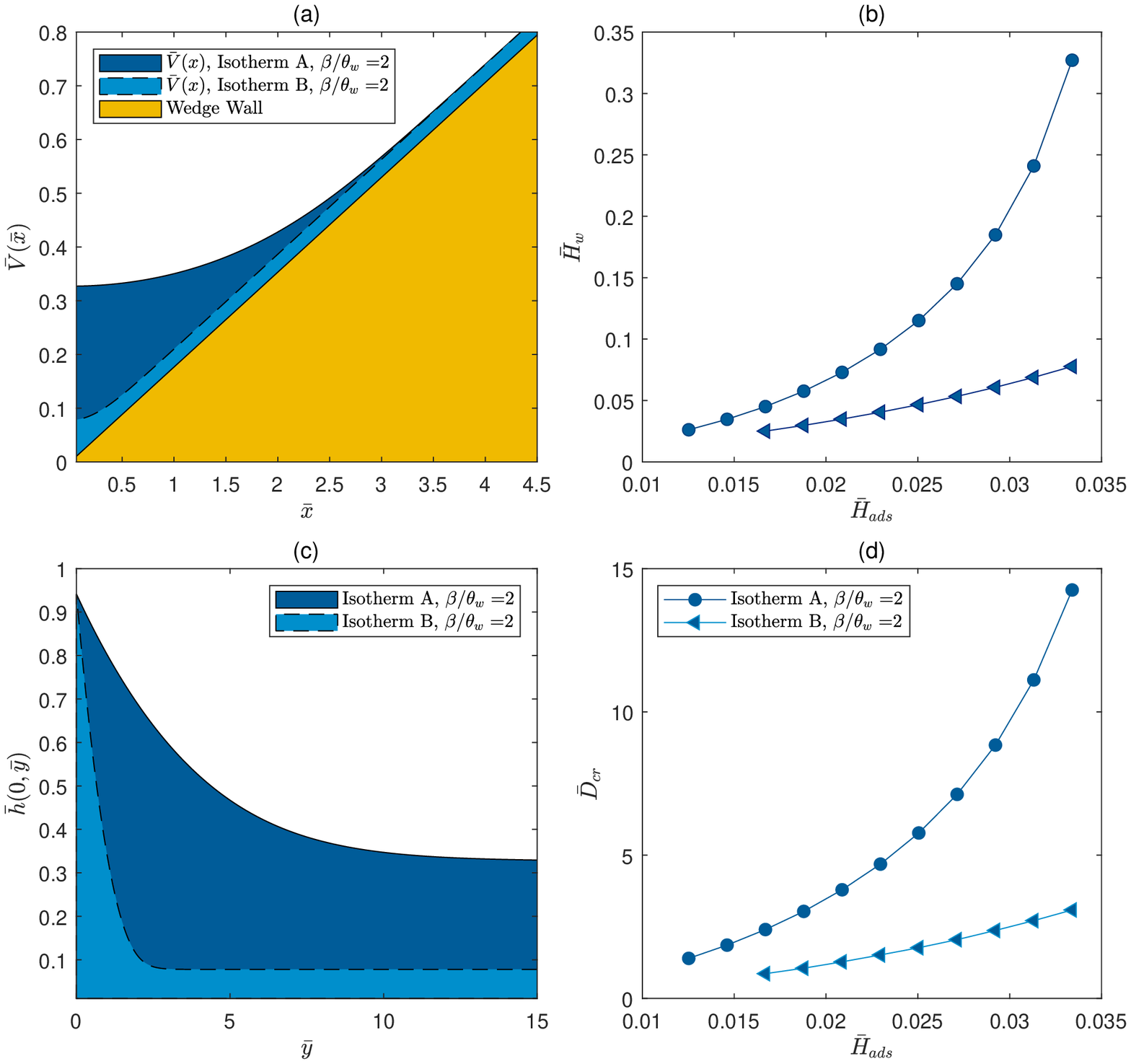}
\caption{Influence of the disjoining pressure on the wetting of the wedge; (a) limiting profiles $\bar V(\bar x)$ of the rivulets calculated with the adsorbed film thicknesses $\bar H_{ads} = 0.0334$; (b) dependence of the limiting height $\bar H_{w}$ on the  adsorbed film thickness; (c) thickness of the rivulet at the symmetry axis ($\bar x = 0$) calculated with the adsorbed film thicknesses $\bar H_{ads} = 0.0334$; (d) dependence of the decay length $\bar D_{cr}$ on the adsorbed film thickness. The inclination angle is $\beta = 2\theta_{w}$. The lines in (b) and (d) are to guide an eye.}
\label{F_Influence_Isotherm}
\end{figure}

As can be expected, the shape of the rivulet will be different for different liquid systems. It is, however, of substantial interest how the surface forces affect the steady state of nanorivulets in the systems having the same macroscopic wetting behavior (same $\theta_w$ and $\mathfrak{\bar S}(\bar H_{1})$). In the following, we show how the disjoining pressure isotherm influences the shape of the meniscus in the wedge. 

Two isotherms shown in Figure \ref{F_Isotherms} have been employed. We note that dimensional values of $H_{ads}$ for isotherm B were smaller than $1 \si{\nm}$. However, evaluation of the surface tension according to \cite{Israelachvili1974} accounting for the hydrogen bonds of water \cite{David2014} shows that its alteration does not exceed $1.5\%$. Therefore, the assumption of the constant surface tension is still valid. 

In Figure \ref{F_Influence_Isotherm}(a), the limiting profiles $\bar V(\bar x)$ are shown for both isotherms. Despite the fact that the contact angle $\theta_w$ is preserved, at the nanoscale, the limiting profiles demonstrate significant differences in the limiting heights $\bar H_{w}$. That is caused by the interplay of long-range and short-range surface forces. Since wedges of the same $\bar X_w$, $\bar Y_w$, $\bar H_{ads}$ and $\beta$ are under consideration, limiting height $\bar H_{w}$ is the main parameter impacting the decay length (Figure \ref{F_Influence_Isotherm}, b). The influence of the disjoining pressure isotherm on $\bar H_w$ can be understood from the simplified consideration based on  (\ref{E:Linearized_Equation_Limiting_Height}). Since $\bar H_{ads}$ are the chosen to be the same for both wedges, the differences in the limiting height $\bar H_w$ are mainly rendered by the value of $\bar \Pi(\bar H_{ads})$ which is larger for the isotherm B. The latter causes smaller $\bar H_w$. The same can also be comprehended in a more graphical way -- namely, considering the slope of the alpha-branch of the disjoining pressure isotherm (see Supplementary Material). Since the values of $\bar H_{w}$ are much smaller for the case of isotherm B, the rivulet is forced to reach the limiting profile very close to the inlet (Figure \ref{F_Influence_Isotherm}, c) and has smaller values of $\bar D_{cr}$, as is shown in Figure \ref{F_Influence_Isotherm}(d). Thus, tuning the parameters of the disjoining pressure isotherm can be a way for control of the length, over which the rivulets can extend. Similarly to the trend observed for influence of the inclination angles, the differences between $\bar H_{w}$ and $\bar D_{cr}$ for isotherms A and B are larger when the adsorbed film is thicker (humidity is higher). 

\subsection{Soft wedges}

In many cases, surfaces, which are in contact with liquid, can respond to its action by deformation, dissolution, or swelling \cite{Gielok2017}. Understanding of possible bending/folding of solid structures is of utmost importance in different branches of industry such as micro- and nanofabrication \cite{Tanaka1993}. Shrinkage of cement-based materials when drying also lacks understanding, which is paramount for its quality improvement and control \cite{Beltzung2005}. 

\begin{figure}
\centering\includegraphics[width=0.9\linewidth]{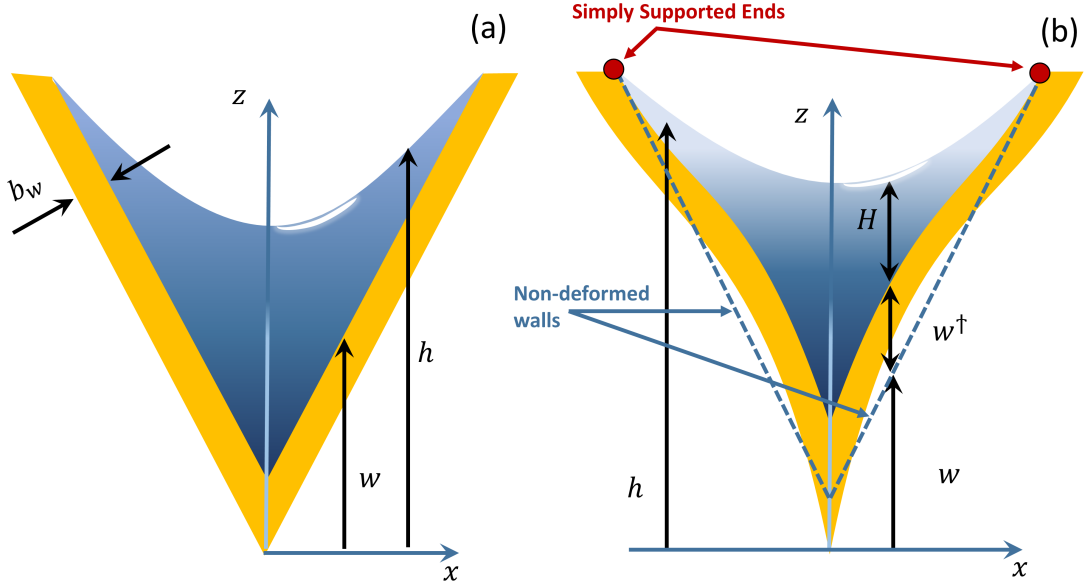}
\caption{Schematic illustration of (a) wedge with rigid walls and (b) wedge with deformable walls. The position of the liquid-gas interface is $h$, the position of the undeformed wall and deformed wall is $w$ and $w^{\dagger}$, accordingly. Thickness of the liquid rivulet filling the wedge is $H$.}
\label{F_Soft_Wedge_Scheme}
\end{figure}

Let us consider the case when a soft wedge is left in the atmosphere of vapor pressure $p_{ads}$, so that an equilibrium adsorbed/wetting film of thickness $H_{ads}$ is formed on its walls. The wedge walls are of constant thickness $b_w$. We assume that the wedge response is purely elastic and instant. It can be defined by its Young modulus $E$ and Poisson ratio $\sigma$. The bending energy of the wedge can be written in the form \cite{Reddy2017}

\begin{equation}
        \mathscr{E}[w^{\dagger}(x,y)] = \iint \left ( D \left (\nabla^2 w^{\dagger} \right)^2  + T w^{\dagger} \right ) dxdy,
    \label{E_Bending_Hamiltonian}
\end{equation}

\noindent where the first term is similar to Euler-Bernoulli elastica with $w^{\dagger}$ being a deflection of the wall of the wedge, and a constant $D$ being the flexural rigidity, $D = \cfrac{E b_w^3}{12 (1 - \sigma^2)}$. The second term is a Lagrange multiplier accounting for a capillary pressure-induced traction $T = \gamma \nabla^2 h$ applied to the wall of the wedge. Minimization of (\ref{E_Bending_Hamiltonian}) leads to the biharmonic Kirchhoff-Love equation defining the steady-state of the wedge \cite{Reddy2017}:

\begin{equation}
     D \nabla^2 \nabla^2 w^{\dagger}  - T = 0. 
    \label{E_Bending_Equation_3D}
\end{equation}

\noindent Since the wedge in the following consideration is not connected to any liquid pool, the liquid profile along the wedge is translationally invariant: $w^{\dagger} = w^{\dagger}(\bar x)$. Therefore, equation  (\ref{E_Bending_Equation_3D}) is reduced to

\begin{equation}
     D \frac{d^4 w^{\dagger}}{dx^4}  - T = 0. 
    \label{E_Bending_Equation_2D}
\end{equation}

\noindent Equation (\ref{E_Bending_Equation_2D}), albeit with a different traction, has been recently used by \citet{Svetovoy2017} to model an adhered cantilever bent by the van der Waals/Casimir forces. In our case of the soft wedge, the traction can be only obtained from the solution of the Derjaguin equation accounting for the possible deformation of the walls. The position of the liquid-gas interface is defined as $\bar h = \bar H + \bar w + \bar w^{\dagger}$ (Figure \ref{F_Soft_Wedge_Scheme}) and, therefore, equation reads

\begin{equation}
     \frac{d^2 \bar H}{d \bar x^2} = \bar\Pi\left (\bar H_{ads} \right ) - \bar\Pi(\bar H) - \frac{d^2 \bar w^{\dagger}}{d \bar x^2},
     \label{E_Derjaguin_2D_soft}
\end{equation}

\noindent Equation (\ref{E_Derjaguin_2D_soft}) is solved as a boundary value problem. In the case of the rigid wedge, the numerical solution can be controlled by comparing with analytical solution (\ref{E_Derjaguin-Solution}). 

To write equation (\ref{E_Bending_Equation_2D}) in a dimensionless form, we use the scaling related to the range of the surface force action $H_2$ used in subsection \ref{SB:Steady_states_of_rivulets}. The deformation of the wedge wall and the wedge's wall thickness reads $\bar w^{\dagger} = \cfrac{w^{\dagger}}{H_2}$ and $\bar b_w = \cfrac{b_w}{H_2}$, accordingly. Defining the dimensionless elastocapillary length $\mathcal{\bar L_C} =\gamma \cfrac{1 - \nu^2}{E H_2}$, we can write the dimensionless traction exerted onto the wall by the liquid as $\bar T = {\mathcal{\bar L_C}}\cfrac{d^2 \bar h}{d\bar x^2}$. Using equation (\ref{E_Derjaguin_2D_soft}), the traction $\bar T$ can also be written as $\bar T = {\mathcal{\bar L_C}}\left [ \bar\Pi\left (\bar H_{ads} \right ) - \bar\Pi(\bar H) \right ]$. Thus, equation  (\ref{E_Bending_Equation_2D}) can be rewritten using dimensionless variables as

\begin{equation}
     \frac{d^4 \bar w^{\dagger}}{d \bar x^4}  - \mathcal{S}\cfrac{d^2 \bar h}{d\bar x^2} = 0, 
    \label{E_Bending_Equation_2D_dimensionless}
\end{equation}

\noindent where $\mathcal{S} = \cfrac{12 \mathcal{\bar L_C}}{\bar b_w^3} = \cfrac{12 \gamma (1 - \sigma^2) H_2^2}{E b_w^3}$ is a softness parameter. The equation is solved as a boundary value problem. The boundary conditions at the symmetry axis are $\cfrac{d \bar w^{\dagger}}{d \bar x}\biggr\vert_{\bar x = 0} = 0$ and $\cfrac{d^3 \bar w^{\dagger}}{d \bar x^3}\biggr\vert_{\bar x = 0} = 0$. The first boundary condition is due to the symmetry. The second boundary condition stands for the absence of a shear force at the symmetry axis. The right end of the wedge is considered to be simply supported and, therefore, we impose the following boundary conditions: $\bar w^{\dagger}(\bar X_w) = 0$ and $\cfrac{d^2 \bar w^{\dagger}}{d \bar x^2}\biggr\vert_{\bar x = \bar X_w} = 0$. The latter condition corresponds to the absence of the bending moment.

\begin{figure}
\centering\includegraphics[width=1\linewidth]{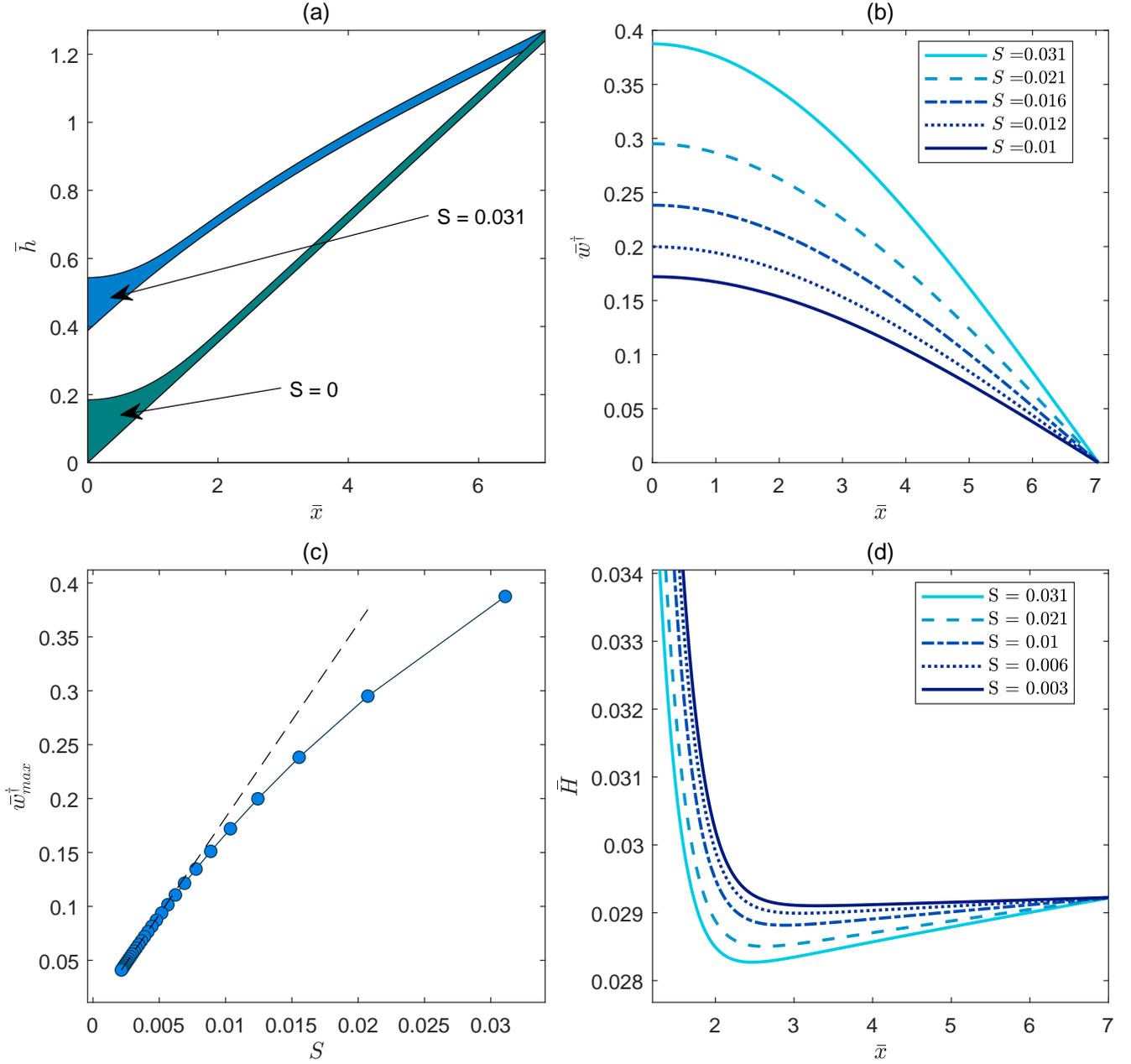}
\caption{Wetting of the wedges with soft walls: (a) Rivulet profiles for soft ($\mathcal{S} = 0.031$) and rigid ($\mathcal{S} = 0$) walls; (b) Deflection of the wedge's wall for different values of softness parameter $\mathcal{S}$; (c)  Dependence of the maximal deformation $\bar w^{\dagger}_{max}$ on the softness parameter $\mathcal{S}$; (d) Thickness of the rivulet in the deformed wedges. The system behavior is defined by the isotherm A, $\beta = 2\theta_{w}$, $\phi = 0.7$.}
\label{F_Soft_Wedges_Diff_S}
\end{figure}

\begin{figure}
\centering\includegraphics[width=1\linewidth]{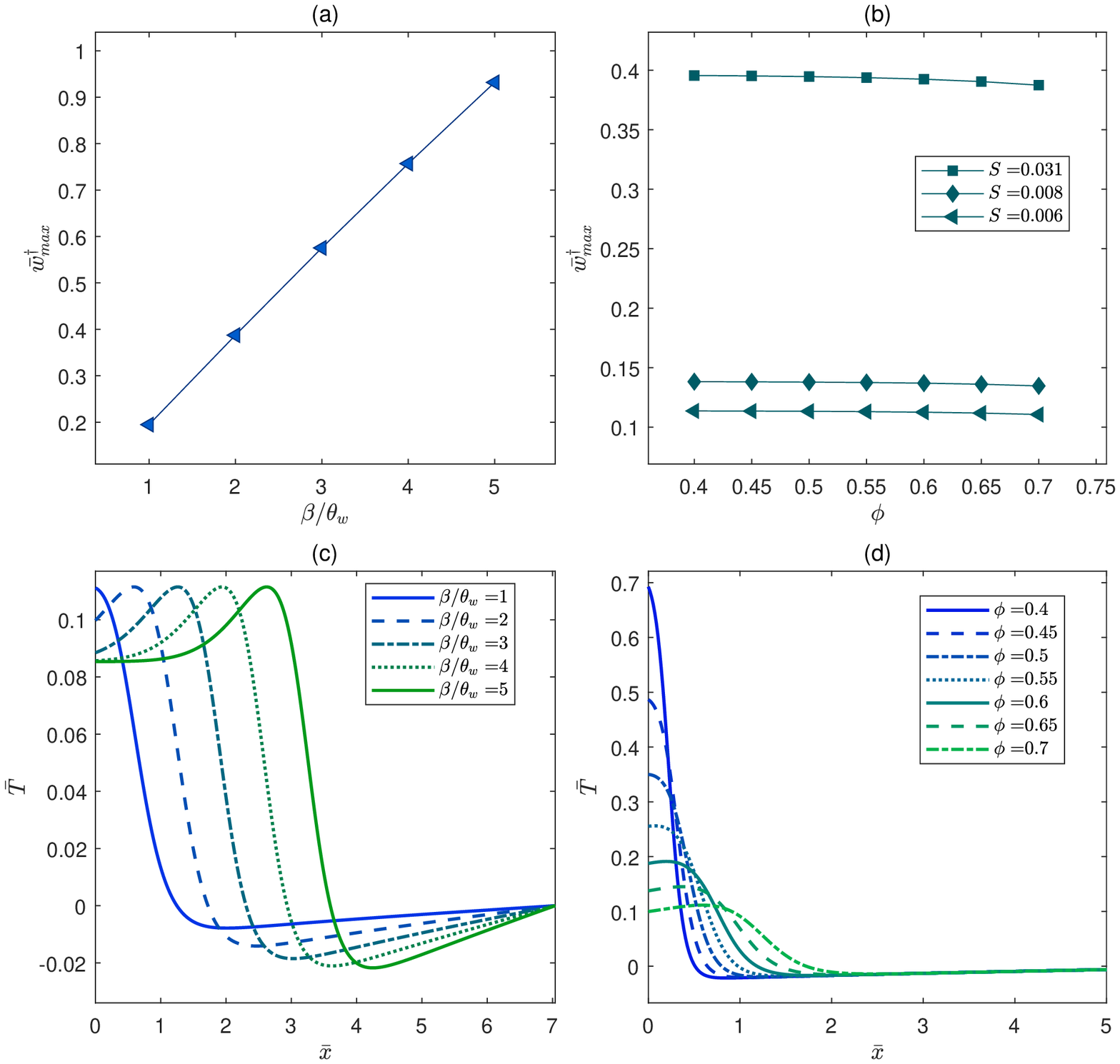}
\caption{Wetting of the wedges with the soft walls. Dependence of the maximal deformation $\bar w^{\dagger}_{max}$ on the (a) inclination angle ($\mathcal{S} = 0.031$, $\phi = 0.7$), (b) on the wetting film thickness (parameter $\phi$) ($\mathcal{S} = 0.031$, $\beta = 2\theta_w$); (c) Traction exerted onto the walls of the wedge inclined by different $\beta$ ($\mathcal{S} = 0.031$, $\phi = 0.7$); (d) Traction exerted onto the walls of the wedge covered by the wetting film of different thickness ($\mathcal{S} = 0.031, \beta = 2\theta_{w}$). The system behavior is defined by the isotherm A.}
\label{F_Soft_Wedges_Param}
\end{figure}

We note that in order to predict the large deformations, equation (\ref{E_Bending_Equation_2D_dimensionless}) has to be solved in iterations, so that the deformation corresponds to the traction induced by the liquid layer covering the deflected wall. Generally, 100 iterations were enough to reach a good convergence of the traction. The solutions of both equations (\ref{E_Derjaguin_2D_soft}) and (\ref{E_Bending_Equation_2D_dimensionless}) have been checked directly by numerical differentiation. The results have been in good agreement with those obtained by the numerical differentiation. The solver for obtaining the deflection has also been checked by comparison with the corresponding analytical solution of (\ref{E_Bending_Equation_2D_dimensionless})

\begin{equation}
     \bar w^{\dagger} = \frac{\mathcal{S} \bar T}{24} \left(5\bar X_w^4  - 6\bar X_w^2 \bar x^2 + \bar x^4 \right ), 
    \label{E_Bending_Analytical_Solution}
\end{equation}

\noindent valid in the case when the traction $\bar T$ is uniformly distributed over the wall of the wedge.

In Figure \ref{F_Soft_Wedges_Diff_S}(a), the effect of the softness parameter on the shape of the rivulet is demonstrated. When the wedge walls are soft, the traction resulted from the interplay of the curvature-induced pressure and disjoining pressure deflects them and pull the wedge tip up.  The deformation of the wedge's wall $\bar w^{\dagger}$ is shown in Figure \ref{F_Soft_Wedges_Diff_S}(b). As can be expected, the deformation becomes more pronounced for larger $\mathcal{S}$ and is maximal at the symmetry axis, since the wedge ends are supported. The dependence of the maximal deformation of the wedge $\bar w^{\dagger}_{max} = \sup_{\bar x \in [0, \bar X_w]}\bar w^{\dagger}$  on the softness parameter $\mathcal{S}$ is illustrated in Figure \ref{F_Soft_Wedges_Diff_S}(c). It can be seen that the dependence is nearly linear for small $\mathcal{S}$ (the linear trend is shown by a dashed black line). The changes of $\bar w^{\dagger}_{max}$ depart from the linear trend for the softer wedges ($\mathcal{S} > 0.05$). Bending the wall of the wedge renders the region with $\bar H < \bar H_{ads}$ (Figure \ref{F_Soft_Wedges_Diff_S}, d), on which the negative traction is exerted. The departure from the linear trend is, hence, related to the shape of the function $\mathcal{S} \bar T$. 

The influence of the wedge inclination angle $\beta$ and wetting film thickness $\bar H_{ads}$ (corresponding parameter $\phi$) on the maximal deformation and traction distribution $\bar T$ is illustrated in Figure \ref{F_Soft_Wedges_Param} (a-d).
The increasing inclination angle $\beta$ leads to the increasing deflection at the symmetry axis of the wedge. As can be seen in Figure \ref{F_Soft_Wedges_Param}(a), this dependence is linear. However, the effect of the inclination angle does not manifest itself via only the limiting thickness $\bar H_{w}$, as could be suggested. The limiting thickness $\bar H_{w}$ increases with the increasing $\beta$. For $\beta/\theta_{w} > 2$, the limiting height is $\bar H_{w} \in \left (\bar H_{c1}; 1 \right )$ (see schematic illustration in Supplementary Materials). Therefore, such increase of $\bar H_{w}$ produces smaller traction at the symmetry axis (Figure \ref{F_Soft_Wedges_Param}, c). The maximal traction in those cases is $\bar \Pi(\bar H_{c1})$. The coordinate $\bar x$, where it is exerted onto the wall, is farther from the symmetry axis for the larger $\bar H_{w}$. That renders wider distributed (spread) traction and, hence, larger deformation, even though the the traction at the symmetry axis is smaller. For the case when $\beta/\theta_{w} = 1$, the limiting height $\bar H_{w}$ lies very close to $\bar H_{c1}$, where the disjoining pressure isotherm finds its minimal value. Therefore, the maximal value of the traction is reached already at the symmetry axis which, in turn, renders a less spread traction.

The influence of the adsorbed/wetting film thickness (parameter $\phi$) on the deformation $\bar w^{\dagger}_{max}$ is presented in Figure \ref{F_Soft_Wedges_Param}(b). Since the formation of the film of thickness $\bar H_{ads}$ is induced by the adsorption onto the walls of the wedge, we also plot a dependence of the maximal deformation of the wall of the wedge on the relative humidity $RH$ of the atmosphere (Supplementary Materials). Unexpectedly, increasing humidity and, therefore, the adsorbed/wetting film thickness leads to the barely observable changes in the maximal deflection. The same trend was obtained for different values of $\mathcal{S}$. One can see from Figure \ref{F_Soft_Wedges_Param}(d) that for smaller $\bar H_{ads}$, the traction has a bigger amplitude $\bar T_{max}$, although it acts mainly in the vicinity of the symmetry axis. In turn, when $\bar H_{ads}$ is larger, the traction has smaller $\bar T_{max}$ albeit wider region of action. Thus, the negligibly small changes of $\bar w^{\dagger}_{max}$ can result from the interplay of the traction amplitude and distribution. 

\begin{figure}[h]
\centering\includegraphics[width=1\linewidth]{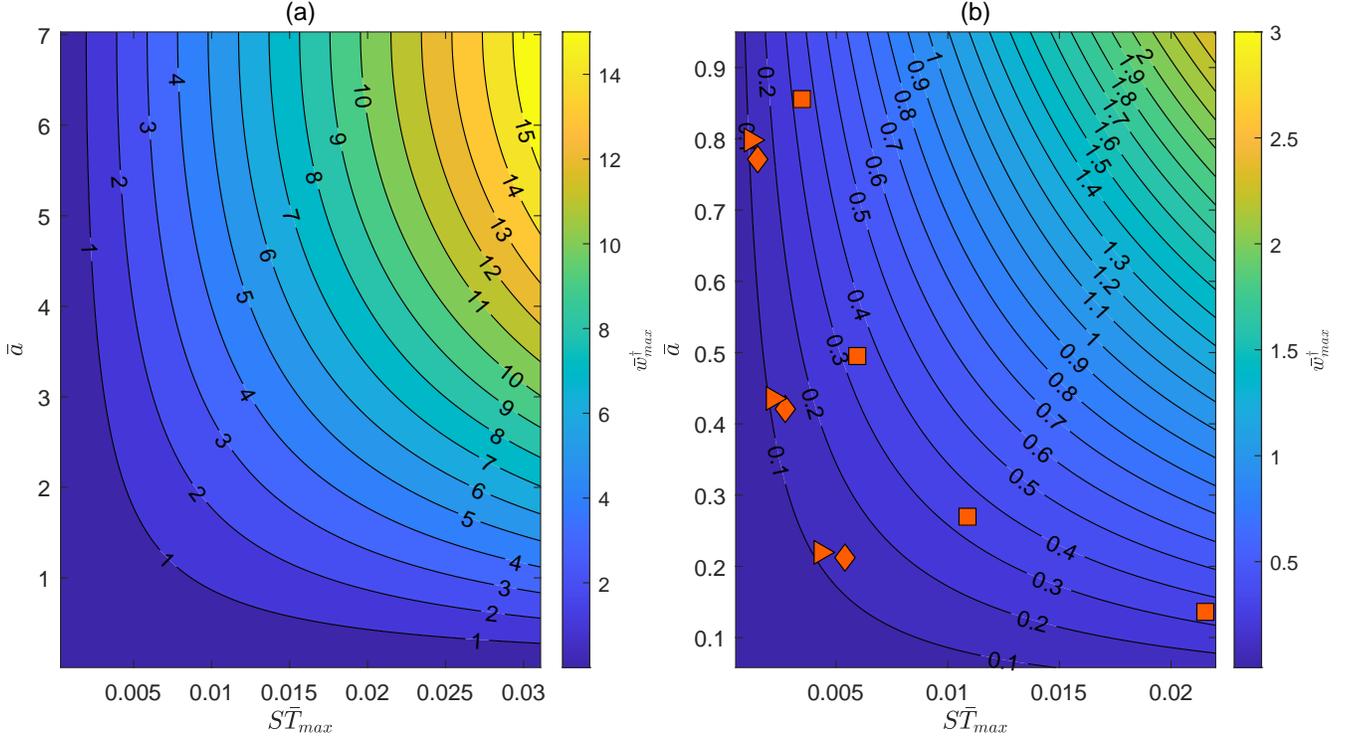}
\caption{(a) The $\mathcal{S}\bar T_{max}$-$\bar a$-map of deformations obtained from solution (\ref{E_Heaviside_Solution}); (b) Magnified region of the map for the small deformations.  The analytical solutions for the cases with $\phi = 0.4, 0.5, 0.6, 0.7$ ($\mathcal{S} = 0.03$), $\phi = 0.4, 0.5, 0.6$ ($\mathcal{S} = 0.008$) and $\phi = 0.4, 0.5, 0.6$ ($\mathcal{S} = 0.006$) are shown by the orange squares, diamonds and triangles, accordingly. They can be compared to the numerically obtained deformations presented in Figure \ref{F_Soft_Wedges_Param}(b).}
\label{F_Soft_Wedges_Map}
\end{figure}

The interplay of breadth of the region $\bar a$, over which the traction $\bar T$ is distributed, and the traction amplitude $\bar T_{max}$ can be better understood from the simpler case suggested by \citet{Svetovoy2017}. The traction can be modeled as a Heaviside function $\bar T_{max}\Theta(\bar a- \bar x)$. Solving the corresponding equation $\cfrac{d^4 \bar w^{\dagger}}{d \bar x^4}  - \mathcal{S}\bar T_{max} \Theta(\bar a- \bar x) = 0$, one can arrive at the deformation at the symmetry axis (see Supplementary Material for the solution):

\begin{equation}
    \bar w^{\dagger}_{max} = \cfrac{\mathcal{S} \bar T_{max}}{24}\left [5\bar X_w^4 + (\bar a - \bar X_w)^4 - 6\bar X_w^2 (\bar a - \bar X_w)^2  \right].
    \label{E_Heaviside_Solution}
\end{equation}

\noindent The calculated deformations in the form of a $\mathcal{S}\bar T_{max}$-$\bar a$-map are shown in Figure \ref{F_Soft_Wedges_Map}(a, b). One can see that a concentrated traction with higher amplitude can still cause the same deformation as a spread traction having much smaller amplitude. In Figure \ref{F_Soft_Wedges_Map}(b), the analytical solutions for the cases with $\phi = 0.4, 0.5, 0.6, 0.7$ are shown by orange squares and triangles for $\mathcal{S} = 0.03$ and $\mathcal{S} = 0.006$, accordingly. The value of breadth $\bar a$ cannot be obtained analytically. It has been evaluated using the numerical results as $\bar a = \cfrac{\int_0^{\bar X_w} \bar T d\bar x}{\sup_{\bar x \in [0, \bar X_w]}\bar T}$, so the forces per unit length exerted onto the wall of the wedge by the real traction (obtained numerically) and the Heaviside traction are equal. Although the deformations obtained analytically slightly depart from the numerical values (Figure \ref{F_Soft_Wedges_Param}, b), they generally support the findings, being located along the lines of the equal deformations.

\section{Conclusions}

In the present work, we model edge wetting and discuss the shape of rivulets in rigid and soft open wedge channels. We show that when intermolecular interactions (surface forces) are taken into account, the validity of the well-known Concus-Finn condition breaks down. The steady-state rivulets can exist in the wedges even when the liquid-gas interface is concave. In order to obtain and quantify the rivulet profiles, we use the disjoining pressure concept and solve the Derjaguin equation. We demonstrate that the extent of the rivulet drastically depends on the humidity of the surrounding atmosphere, wedge geometry and surface force parameters. 

We use Kirchhoff-Love theory in order to model the deformation of the soft wedge induced by the presence of the rivulet. We show that the relative humidity does not affect the deflection of the walls of the wedge, whereas the geometry of the wedge does. 

We present simple analytical models allowing to predict the limiting height of the rivulet in the wedge and to estimate the deformation resulting from the interplay of the amplitude of the traction exerted onto the walls of the wedge and the breadth of the region, over which the traction is distributed. \newline

\noindent \textbf{Acknowledgements} \\
The authors gratefully acknowledge the financial support from the Deutsche Forschungsgemeinschaft (DFG, German Research Foundation) – Priority Program “Dynamic Wetting of Flexible, Adaptive and Switchable Surfaces” (SPP 2171), Project Number 422792679.

\section*{Supplementary Materials}

\subsection*{Disjoining pressure isotherm}

The advantage of using the disjoining pressure isotherm containing a dispersive power-law part $\cfrac{A}{H^n}$ ($n$ defines the dispersive repulsion/attraction) and a weak-overlap electrostatic/structural exponential tail $Ke^{-H/\chi}$ is the easier evaluation of $H_1$, $H_2$ as well as its minima and maxima. The function $\Pi(H)$ is differentiable for $H \in (0; \infty)$ with the derivatives 

\begin{equation}
    \Pi^{(k)}(H) = (-1)^k \left (A\frac{ \prod_{i=0}^{i = k-1} (n+ i)}{H^{n + k}} - K\frac{e^{-H/\chi}}{\chi^k} \right),
\end{equation}

\noindent where $k$ is the order of the derivative. For the van der Waals forces, we have $n=3$, whereas for the Casimir forces, we have $n=4$. The roots of equation $ \Pi^{(k)}(H) = 0$, that is

\begin{equation}
     A\frac{\prod_{i=0}^{i = k-1} (n+ i) }{H^{n + k}} - K\frac{e^{-H/\chi}}{\chi^k} = 0,
\end{equation}

\noindent where $k \geq 1$, can be found using Lambert $\mathcal{W}$-function with branches $l=0$ and $l=-1$, read as 

\begin{equation}
    H_{c\{k, l\}} = -(n+k) \chi \mathcal{W}_l \left[ - \frac{1}{(n+k) \chi}\left( \frac{A \chi^k  \prod_{i=0}^{i = k-1} (n+ i)}{K} \right )^{\frac{1}{(n+k)}}\right ]
    \label{roots_DP_deriv}
\end{equation}

\noindent For the case when $k=0$, we obtain zeros of the disjoining pressure isotherm

\begin{equation}
    H_{c\{0,0\}} = H_{1} = -n\chi  \mathcal{W}_0 \left[ - \frac{1}{n \chi}\left( \frac{A}{K} \right )^{\frac{1}{n}}\right ]
\end{equation}

\begin{equation}
    H_{c\{0,-1\}} = H_{2} = -n\chi  \mathcal{W}_{-1} \left[ - \frac{1}{n \chi}\left( \frac{A}{K} \right )^{\frac{1}{n}}\right ]
\end{equation}

\noindent When $k=1$, zeros of the derivative of the disjoining pressure isotherm read

\begin{equation}
    H_{c\{1, 0\}} = H_{c1} = -(n+1) \chi \mathcal{W}_0 \left[ - \frac{1}{(n+1) \chi}\left( \frac{nA \chi}{K} \right )^{\frac{1}{(n+1)}}\right ]
    \label{roots_DP_deriv_11}
\end{equation}

\begin{equation}
    H_{c\{1, -1\}} = H_{c2} = -(n+1) \chi \mathcal{W}_{-1} \left[ - \frac{1}{(n+1) \chi}\left( \frac{nA \chi}{K} \right )^{\frac{1}{(n+1)}}\right ]
    \label{roots_DP_deriv_12}
\end{equation}

\subsection*{Inlet profile construction}

The inlet profile is assumed to be macroscopic. Thus, it results from the solution of the Young-Laplace equation:

\begin{equation}
    \frac{d^2 \bar h(\bar x)}{d \bar x^2} = \bar p_c,
    \label{E:YL_eq}
\end{equation}

\noindent where $\bar p_c$ is the capillary pressure. The  boundary conditions are $\cfrac{d \bar h}{d \bar x} = 0$ at $\bar x = 0$, $\bar h = \bar H_{ads} + \bar X_w \tan\beta$ at $\bar x = \bar X_w$. Integrating (\ref{E:YL_eq}), one arrives at

\begin{equation}
    \bar h(\bar x) = \bar p_c \frac{\left ( \bar x^2 - \bar X_w^2 \right )}{2} + \bar X_w \tan\beta + \bar H_{ads}.
    \label{E:YL_eq_sol}
\end{equation}

\begin{figure}[h]
\centering\includegraphics[width=0.8\linewidth]{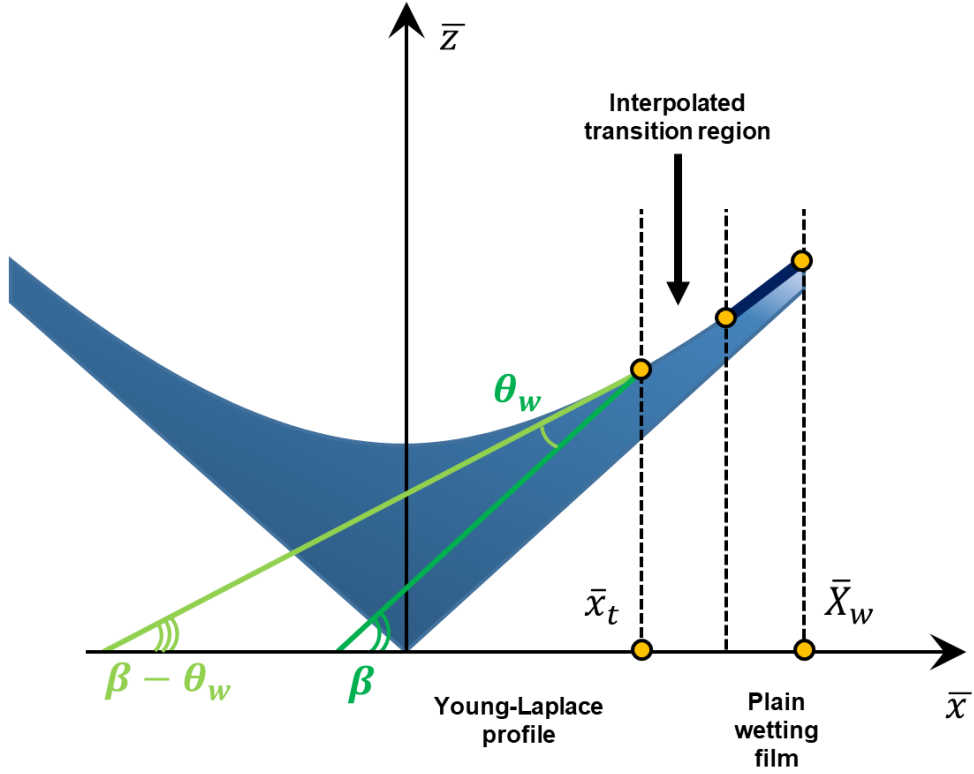}
\caption{Schematic representation of the inlet profile.
}
\label{F_Inlet}
\end{figure}

\noindent This solution, however, does not satisfy the boundary condition for the Derjaguin equation for the rivulet, since it does not contain wetting film, to which the interface relaxes (Figure \ref{F_Inlet}). Therefore, we must introduce a point $x_t$, where the transition to the wetting film occurs. Since the macroscopic profile does not immeadiately reaches the wetting film, at the transition point $\bar x_t$ it has the thickness $\bar H_{ads} + c_1 \bar H_{ads}$, where $c_1$ is a constant. Equation (\ref{E:YL_eq_sol}) will be rewritten as

\begin{equation}
    \bar h(\bar x) = \bar p_c \frac{\left ( \bar x^2 - \bar x_t^2 \right )}{2} + \bar x_t \tan\beta + \bar H_{ads} + c_1 \bar H_{ads}.
    \label{E:YL_eq_sol_2}
\end{equation}

\noindent The value of the capillary pressure can be found from the condition on the contact angle $\theta_w$:  $\cfrac{d \bar h}{d \bar x} = \tan(\beta - \theta_w)$ at $\bar x = \bar x_t$. The capillary pressure is obtained as $\bar p_c = \cfrac{\tan(\beta - \theta_w)}{\bar  x_t}$, and the shape of the interface for $\bar x \leq \bar x_t$ is written 

\begin{equation}
    \bar h(\bar x) = \frac{\tan(\beta - \theta_w)}{\bar x_t} \frac{\left (\bar x^2 - \bar x_t^2 \right )}{2} + \bar x_t \tan\beta + \bar H_{ads} + c_1 \bar H_{ads}.
    \label{E:YL_eq_sol_3}
\end{equation}

\noindent In order to make the transition smooth, we assume that the interface is the plane wetting film when $\bar x \geq \bar x_t + c_2\left (\bar X_w - \bar x_t \right)$ and interpolate the region $\bar x \in \left(\bar x_t; \bar x_t + c_2\left (\bar X_w - \bar x_t \right)\right)$.

The parameters $\bar x_t$, $c_1$, and $c_2$ of the inlet profile \ref{E:YL_eq_sol_3} used for the simulations are presented in Table \ref{TS1}. Slight variation of the parameters did not lead to significant changes in $\bar H(0, \bar y)$.

\begin{table}[h]

\centering
\begin{tabular}{l l l l l l l l l}
\hline
\\
$\bar x_t$ & $c_1$ & $c_2$ & Geometry
\\
\\
\hline
Data in Figures 3-5, 7, 10
\\
\hline
$0.95\bar X_w$  & $0.6$ & $0.650$ &  All configurations presented\\ 
\hline
Data in Figure 8
\\
\hline
$0.950\bar X_w$  & $0.6$ & $0.650$  & $\bar X_w = 7.0$, $\beta = \theta$, $\phi = 0.3$ \\ 
$0.631\bar X_w$ & $0.3$ & $0.030$ & $\bar X_w = 7.0$, $\beta = 2\theta$, $\phi = 0.3$ \\
$0.465\bar X_w$ & $0.3$ & $0.030$ & $\bar X_w = 7.0$, $\beta = 3\theta$, $\phi = 0.3$ \\
$0.364\bar X_w$ & $0.3$ & $0.025$ & $\bar X_w = 7.0$, $\beta = 4\theta$, $\phi = 0.3$ \\
$0.294\bar X_w$ & $0.3$ & $0.015$ & $\bar X_w = 7.0$, $\beta = 5\theta$, $\phi = 0.3$ \\
\hline
Data in Figure 9
\\
\hline
$0.950\bar X_w$  & $0.6$ & $0.650$ & $\bar X_w = 7.0$, $\beta = 2\theta$, $\phi = 0.5$ \\ 
$0.950\bar X_w$ & $0.6$ & $0.600$ & $\bar X_w = 10.6$, $\beta = 2\theta$, $\phi = 0.5$ \\
$0.950\bar X_w$ & $0.6$ & $0.550$ & $\bar X_w = 14.1$, $\beta = 2\theta$, $\phi = 0.5$ \\
$0.950\bar X_w$ & $0.6$ & $0.500$ & $\bar X_w = 17.6$, $\beta = 2\theta$, $\phi = 0.5$ \\
$0.950\bar X_w$ & $0.6$ & $0.450$ & $\bar X_w = 21.1$, $\beta = 2\theta$, $\phi = 0.5$ \\
$0.950\bar X_w$ & $0.6$ & $0.380$ & $\bar X_w = 24.6$, $\beta = 2\theta$, $\phi = 0.5$ \\
$0.950\bar X_w$ & $0.6$ & $0.350$ & $\bar X_w = 28.2$, $\beta = 2\theta$, $\phi = 0.5$ \\
\hline

\end{tabular}
\caption{Parameters of the inlet profile \ref{E:YL_eq_sol_3} used for the simulations.}
\label{TS1}
\end{table}

\subsection*{Limiting height of the rivulet for \\ the linearized disjoining pressure isotherm}

When the wetting film is very thin: $\bar H_{ads} = \phi \bar H_1$ with $\phi \ll 1$, one can expect the limiting height of the rivulet $\bar H_w$ to be smaller than the height $\bar H_{c1}$ (from \ref{roots_DP_deriv_11} with $n=3$) 

\begin{equation}
     H_{c1} = -4 \chi \mathcal{W}_{0} \left[ - \frac{1}{4 \chi^{\frac{3}{4}}}\left( \frac{3 A}{K} \right )^{\frac{1}{4}}\right ],
    \label{roots_1_DP_deriv}
\end{equation}

\noindent where the disjoining pressure isotherm finds its minimum. 

\begin{figure}[h]
\centering\includegraphics[width=1\linewidth]{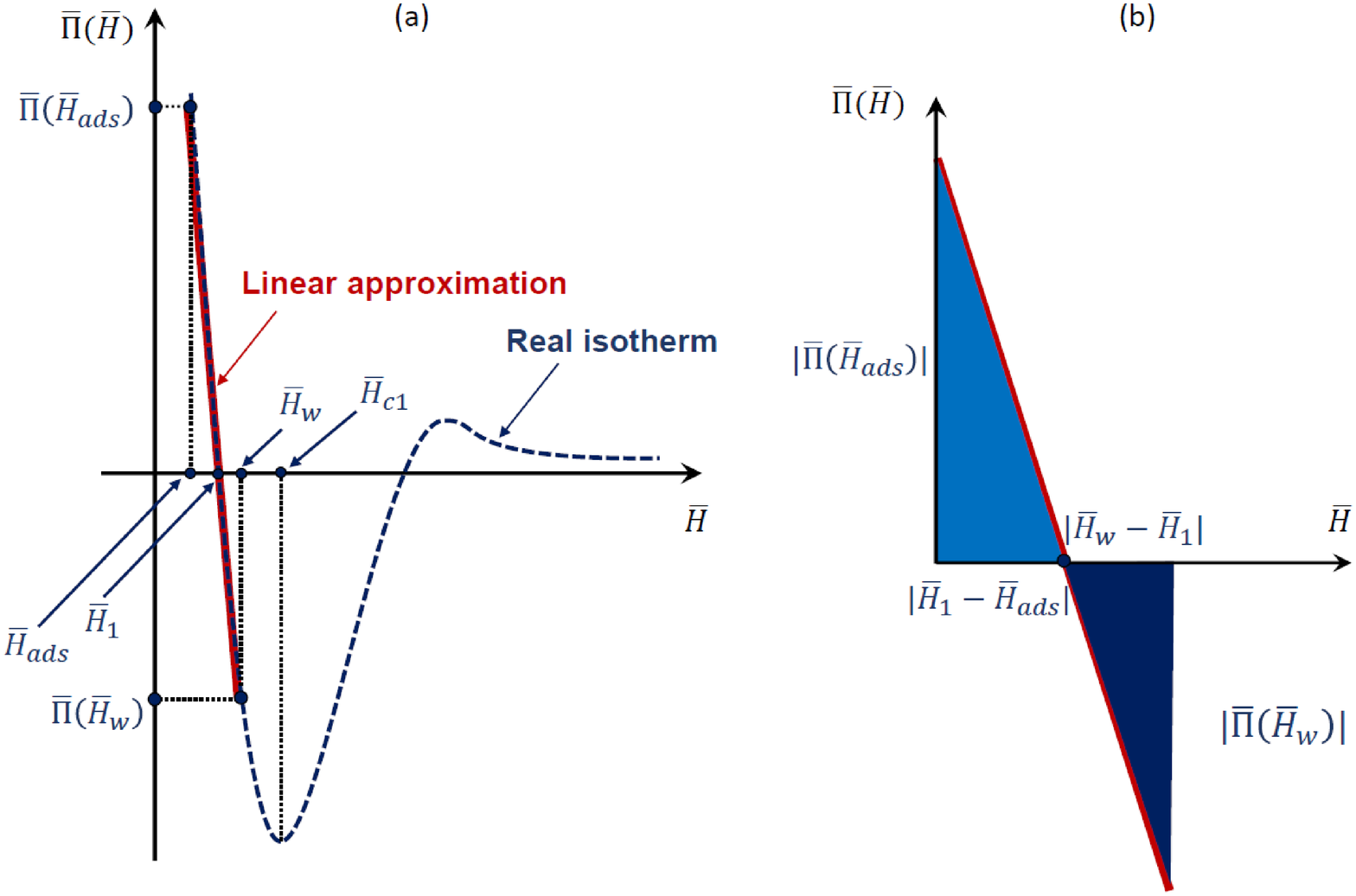}
\caption{ (a) Schematic representation of the disjoining pressure isotherm (dashed dark blue line) and its linearized branch (thick solid red line); (b) region used for the evaluation of integral in (\ref{E_Middle_Height_Supp}).
}
\label{F_Linear_Isotherm}
\end{figure}

\noindent In this case, consideration of only the first stable branch of the disjoining pressure isotherm is sufficient. Due to its shape, this branch can be fairly represented by a straight line connecting points $(\bar H_{ads}; \Pi\left (\bar H_{ads} \right ))$ and $(\bar H_w; \Pi\left (\bar H_w \right ))$ where $\Pi\left (\bar H_w \right )$ (Figure \ref{F_Linear_Isotherm}, a). We also assume that simplified isotherm crosses $\bar H$-axis in the immediate vicinity to $\bar H_1$ so it can be assumed that the point $\left (\bar H_1 , 0\right )$ belongs to this straight line.

\noindent Equation 

\begin{equation}
     \frac{\tan^2\beta}{2} = \bar \Pi\left (\bar H_{ads} \right ) \left (\bar H_{w} -  \bar H_{ads} \right ) - \int_{\bar H_{ads}}^{\bar H_w} \bar\Pi(\bar H) d \bar H
     \label{E_Middle_Height_Supp}
\end{equation}

\noindent can be rewritten in the following form

\begin{equation}
     \tan^2\beta = 2\bar \Pi\left (\bar H_{ads} \right ) \left (\bar H_{w} -  \bar H_{1} \right ) + 2  \Pi\left (\bar H_{ads} \right ) \left (\bar H_{1} -  \bar H_{ads} \right ) - 2\int_{\bar H_{ads}}^{\bar H_w} \bar\Pi(\bar H)  d \bar H,
     \label{E_Middle_Height_Supp2}
\end{equation}

\noindent where the last term can be easily evaluated when the disjoining pressure isotherm is linearized (Figure \ref{F_Linear_Isotherm}, b)

\begin{equation}
     2\int_{\bar H_{ads}}^{\bar H_w} \bar\Pi(\bar H)  d \bar H = \bar \Pi\left (\bar H_{ads} \right ) \left [ \left (\bar H_{1} -  \bar H_{ads} \right ) - \frac{\left (\bar H_{w} -  \bar H_{1} \right)^2}{\bar H_{1} -  \bar H_{ads}} \right].
     \label{E_Middle_Height_Supp3}
\end{equation}

\noindent After some elementary arithmetical operations with (\ref{E_Middle_Height_Supp2}) and (\ref{E_Middle_Height_Supp3}), one obtains

\begin{equation}
    \left (\bar H_{w} -  \bar H_{ads} \right)^2 = \frac{\tan^2\beta \left (\bar H_{1} -  \bar H_{ads} \right ) }{\bar\Pi\left (\bar H_{ads} \right )},
     \label{E_Middle_Height_Supp4}
\end{equation}

\noindent or, introducing the parameter $\phi$,
\begin{equation}
    \bar H_w = \phi \bar H_{1} + \tan\beta \sqrt{\cfrac{\bar H_{1}(1 -  \phi)}{\bar \Pi\left (\phi \bar H_{1} \right )}} = \bar H_{ads} + \tan\beta \sqrt{\zeta^{-2}},
     \label{E_Middle_Height_Supp_Final} 
\end{equation}

\noindent where $-\bar \zeta^2$ is the slope of the first branch of the simplified disjoining pressure isotherm. Thus, among two rivulets having the same wetting film thickness $\bar H_{ads}$, the limiting height $\bar H_w$ can be expected to be smaller for the rivulet, whose disjoining pressure isotherm has a larger slope of the first stable branch (alpha-branch).

\begin{figure}[h]
\centering\includegraphics[width=0.7\linewidth]{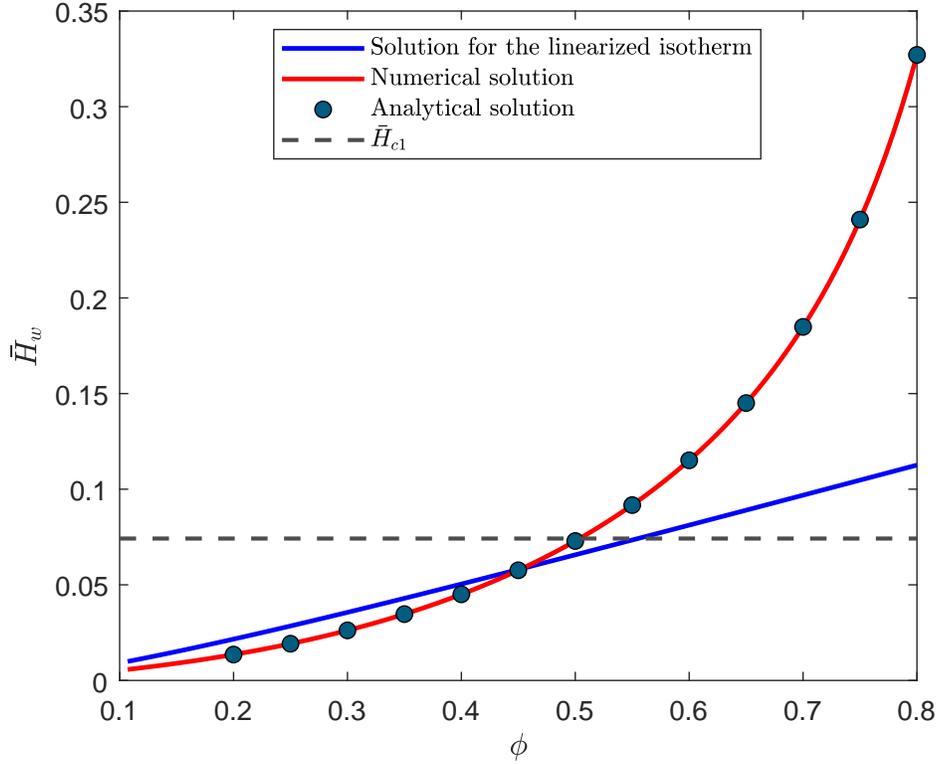}
\caption{Dependence of the limiting height $\bar H_w$ of the rivulet on $\phi$. Blue solid line corresponds to the solution of equation (\ref{E_Middle_Height_Supp_Final}), teal circle markers correspond to $\bar H_w$ obtained from the analytical solution of the Derjaguin equation (as described in the manuscript, Eq. (7) and Eq. (8)), solid red line corresponds to $H_w$ obtained from the numerical solution of the Derjaguin equation (solved as a boundary value problem). The isotherm A has been employed for the calculations.
}
\label{F_Limiting_Height_Supp}
\end{figure}

\noindent The limiting height $\bar H_w$ evaluated using equation (\ref{E_Middle_Height_Supp_Final}) is presented in Figure \ref{F_Limiting_Height_Supp} alongside the numerical results. The dashed black line depicts $\bar H_{c1}$ setting the limiting $\phi$, at which the assumptions used for (\ref{E_Middle_Height_Supp_Final}) are valid. It can be seen that (\ref{E_Middle_Height_Supp_Final}) generally predicts the dependence well for $\phi \lessapprox 0.5$, even despite the fact that the actual shape of the disjoining pressure isotherm has been ignored. We highlight that for the sake of simplicity, we have not accounted for the height dependence of the surface tension, although films with $\phi < 0.25$ are thinner than $1$ nm.

\subsection*{Deformation of the soft wedge: \\ Solution of Kirchhoff-Love equation for a Heaviside traction}

Let us consider the case when a step-like traction exerted onto the substrate near its end, in the region $\bar x \in [0 ; \bar a]$. The traction can be modeled as a Heaviside function $\bar T = \bar T_{max}\Theta(\bar a- \bar x)$ of amplitude $\bar T_{max}$. The solid will respond to the traction by deflecting. The deflection $\bar w^{\dagger}$ satisfies the Kirchhoff-Love equation:

\begin{equation}
    \cfrac{d^4 \bar w^{\dagger}}{d \bar x^4}  - \mathcal{S} \bar T_{max} \Theta(\bar a- \bar x) = 0.
    \label{E:Kirchhoff-Love}
\end{equation}

\noindent We rewrite \ref{E:Kirchhoff-Love} using Laplace transforms. For the fourth derivative, we have 

\begin{equation}
    \mathcal{L}\left [\cfrac{d^4 \bar w^{\dagger}(\bar x)}{d \bar x^4}\right ] = s^4 \mathcal{L}[\bar w(\bar x)]-s^3 \bar w^{\dagger}(0)-s^2 \bar w^{\dagger '}(0)-s \bar w^{\dagger ''}(0)-\bar w^{\dagger(3)}(0),
    \label{E:Deriv_Laplace}
\end{equation}

\noindent where primes denote the derivatives with respect to $s$. From the boundary conditions, we have $\bar w^{\dagger '}(0) = 0$  and $\bar w^{\dagger(3)}(0) = 0$. The Heaviside function can be written in the form

\begin{equation}
    \Theta(\bar a- \bar x)= 1 - \Theta(\bar x - \bar a),
    \label{E:Heaviside}
\end{equation}

\noindent and its Laplace transform reads

\begin{equation}
    \mathcal{L}[1-\Theta (\bar x-\bar a)] = \left(\cfrac{1}{s}-\cfrac{e^{-a s}}{s}\right).
    \label{E:Heaviside_Laplace}
\end{equation}

\noindent Therefore, we arrive at the equation in $s$-domain

\begin{equation}
    \mathcal{L}[\bar w^{\dagger}(\bar x)] = \frac{\mathcal{S}\bar T_{max} \left(1-e^{-a s}\right)}{s^5}+\frac{w''(0)}{s^3}+\frac{w(0)}{s},
\end{equation}

\noindent which can be easily written in the original $\bar x$-domain by the reversed Laplace transform

\begin{equation}
    \bar w^{\dagger}(\bar x) = \frac{1}{24} \mathcal{S}\bar T_{max} \left[\bar x^4 -(\bar a-\bar x)^4 \Theta (\bar x-\bar a)\right] +\frac{\bar x^2}{2}  \bar w^{\dagger''}(0)+\bar w^{\dagger}(0).
\end{equation}

\noindent Due to the presence of the Heaviside function, it is more convenient to consider two intervals:

\begin{equation}
  \bar w^{\dagger}(\bar x) = 
  \begin{cases}
    \cfrac{1}{24} \mathcal{S}\bar T_{max} \left[\bar x^4 -(\bar a-\bar x)^4 \right] +\cfrac{\bar x^2}{2}  \bar w^{\dagger''}(0)+\bar w^{\dagger}(0), & \text{if $\bar x \leq \bar a$} \\
    \cfrac{1}{24} \mathcal{S}\bar T_{max} \bar x^4  +\cfrac{\bar x^2}{2}  \bar w^{\dagger''}(0)+\bar w^{\dagger}(0). &\text{if $\bar x > \bar a$}.
  \end{cases}
\end{equation}

\noindent From the two boundary conditions left $\bar w^{\dagger }(\bar X_w) = 0$  and $\bar w^{\dagger ''}(\bar X_w) = 0$, we obtain 

\begin{equation}
    \bar w^{\dagger}(0) =  -\frac{\mathcal{S}\bar T_{max} \bar X_w^4}{24} + \frac{\mathcal{S}\bar T_{max} \left(\bar a - \bar X_w \right)^4}{24} - \frac{\mathcal{S}\bar T_{max} \bar w^{\dagger''}(0) \bar X_w^2}{2}, 
\end{equation}

\begin{equation}
    \bar w^{\dagger''}(0) =  \frac{\mathcal{S}\bar T_{max} \left (\bar a - \bar X_w \right)^2}{2} - \frac{\mathcal{S}\bar T_{max} \bar X_w^2}{2}.
\end{equation}

\noindent The deformation at the symmetry axis reads

\begin{equation}
    \bar w^{\dagger}(0) =  \frac{\mathcal{S}\bar T_{max}}{24} \left [5 \bar X_w^4 + (\bar a-\bar X_w)^4 - 6\bar X_w^2 (\bar a-\bar X_w)^2 \right]. 
\end{equation}

\begin{figure}
\centering\includegraphics[width=0.9\linewidth]{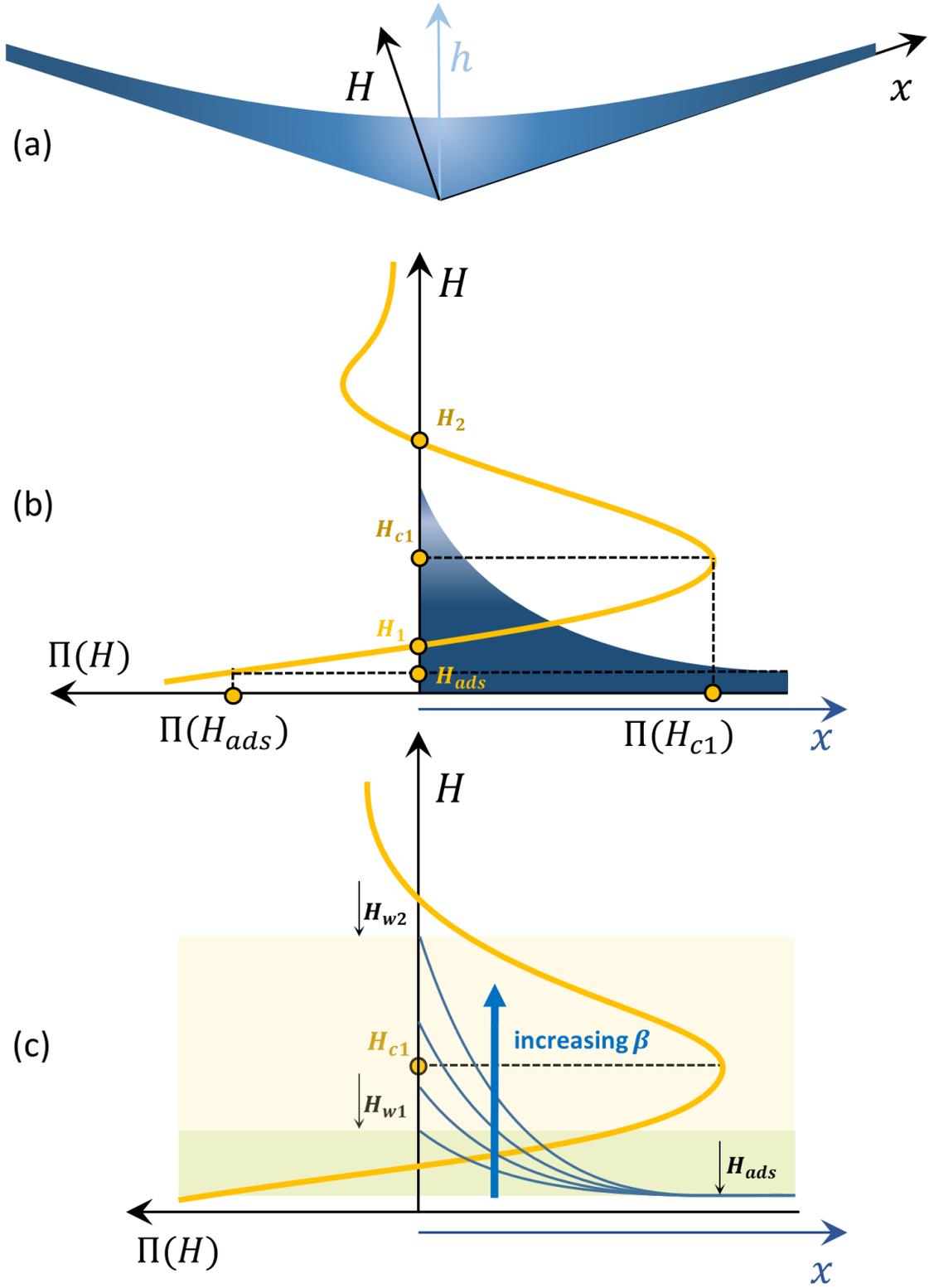}
\caption{(Schematic representation (a) of the wedge; (b) of the distribution of the disjoining pressure over the wedge; (c) influence of the increasing wedge inclination angle on the distribution of the disjoining pressure over the wedge.
}
\label{F_DP_Distribution_Wedge}
\end{figure}

\begin{figure}
\centering\includegraphics[width=0.8\linewidth]{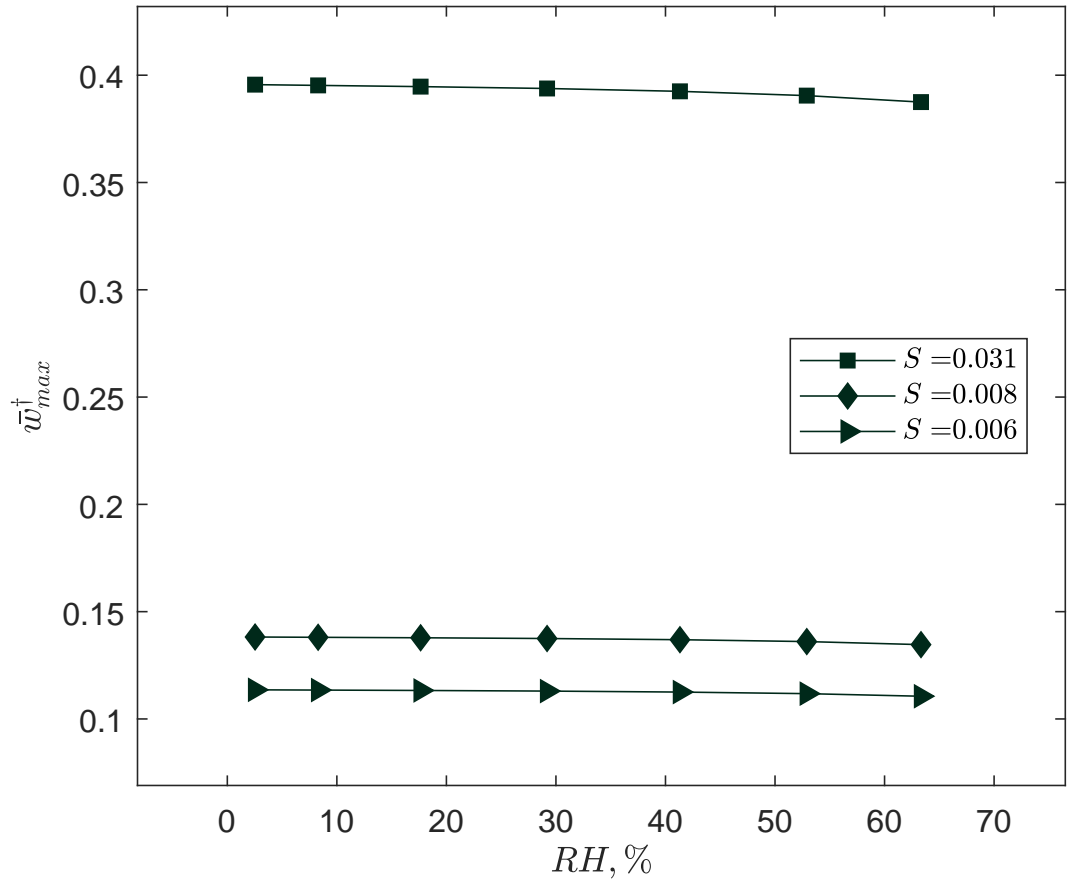}
\caption{Influence of the relative humidity $RH$ on the deformation of the wedge.
}
\label{F_DP_Distribution_Wedge}
\end{figure}

\bibliographystyle{unsrtnat}
\bibliography{sample.bib}

\end{document}